\documentclass[superscriptaddress,aps,twocolumn,longbibliography]{revtex4-1}
\usepackage{amsmath,amssymb}
\usepackage{graphicx}
\usepackage{dcolumn}
\usepackage{enumitem}
\usepackage[normalem]{ulem}
\usepackage{float}
\usepackage{bm,dsfont}
\usepackage{multirow}
\usepackage{simpler-wick}
\usepackage{cuted}
\usepackage{color}
\usepackage{hyperref}
\hypersetup{
	colorlinks = true,
	linkcolor = [rgb]{0.70,0.13,0.13},
	citecolor = [rgb]{0.13,0.55,0.13},
	urlcolor = [rgb]{0.25, 0.41, 0.88}}
\newcommand{\bra}[1]{{\langle #1|}}
\newcommand{\ket}[1]{{|#1 \rangle}}
\newcommand{\braket}[2]{{\langle #1|#2\rangle}}
\newcommand{\Bra}[1]{{\langle\!\langle #1|}}
\newcommand{\Ket}[1]{{|#1 \rangle\!\rangle}}
\newcommand{\Braket}[2]{{\langle\!\langle #1|#2\rangle\!\rangle}}

\newcommand{\id}{\mathds{1}}
\newcommand{\e}{\mathrm{e}}
\newcommand{\U}{\mathrm{U}}

\newcommand{\dsE}{\mathbb{E}}

\newcommand{\scC}{\mathcal{C}}

\newcommand{\scE}{\mathcal{E}}
\newcommand{\scL}{\mathcal{L}}
\newcommand{\scH}{\mathcal{H}}
\newcommand{\scM}{\mathcal{M}}

\newcommand{\scK}{\mathcal{K}}

\newcommand{\scP}{\mathcal{P}}

\newcommand{\scU}{\mathcal{U}}

\newcommand{\Tr}{\mathop{\operatorname{Tr}}}
\newcommand{\E}{\mathop{\dsE}}
\newcommand{\var}{\mathop{\operatorname{var}}}
\newcommand{\supp}{\mathop{\operatorname{supp}}}

\newcommand{\vect}[1]{{\bm{#1}}}

\newcommand{\mat}[1]{\left[\begin{matrix}#1\end{matrix}\right]}

\newcommand{\eq}[1]{\begin{equation}#1\end{equation}}
\newcommand{\eqs}[1]{\begin{equation}\begin{split}#1\end{split}\end{equation}}
\newcommand{\eqnref}[1]{Eq.\,\eqref{#1}}
\newcommand{\figref}[1]{Fig.\,\ref{#1}}

\newcommand{\appref}[1]{Appendix\,\ref{#1}}
\newcommand{\refcite}[1]{Ref.\,\cite{#1}}

\usepackage[mathlines]{lineno}

\begin{document}
\title{Measurement-Induced Criticality is Tomographically Optimal}

\author{Ahmed A. Akhtar}
\affiliation{Department of Physics, University of California San Diego, La Jolla, CA 92093, USA}
\author{Hong-Ye Hu}
\affiliation{Department of Physics, Harvard University, 17 Oxford Street, Cambridge, MA 02138, USA}
\affiliation{Harvard Quantum Initiative, Harvard University, 17 Oxford Street, Cambridge, MA 02138, USA}
\author{Yi-Zhuang You}
\affiliation{Department of Physics, University of California San Diego, La Jolla, CA 92093, USA}

\begin{abstract}
We develop a classical shadow tomography protocol utilizing the randomized measurement scheme based on hybrid quantum circuits, which consist of layers of two-qubit random unitary gates mixed with single-qubit random projective measurements. Unlike conventional protocols that perform all measurements by the end of unitary evolutions, our protocol allows measurements to occur at any spacetime position throughout the quantum evolution. We provide a universal classical post-processing strategy to approximately reconstruct the original quantum state from intermittent measurement outcomes given the corresponding random circuit realizations over repeated experiments. We investigated the sample complexity for estimating different observables at different measurement rates of the hybrid quantum circuits. Our result shows that the sample complexity has an optimal scaling at the critical measurement rate when the hybrid circuit undergoes the measurement-induced transition.
%
%
%
\end{abstract}
	
\pacs{Valid PACS appear here} 

\maketitle

\begin{figure}[htbp]
\begin{center}
\includegraphics[scale=0.65]{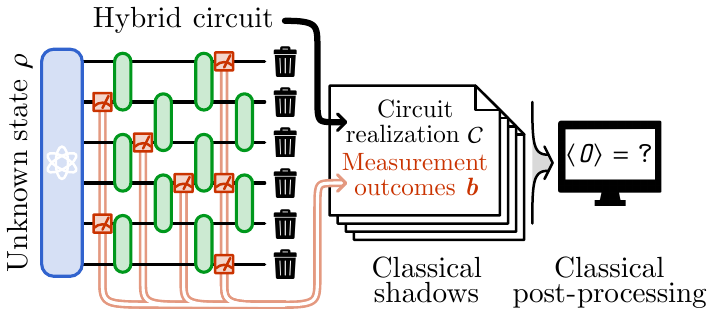}
\caption{Using hybrid quantum circuit as a randomized measurement scheme for classical shadow tomography. Starting from an unknown quantum state $\rho$, evolve the system by layers of random local Clifford gates, and measure each qubit with probability $p$ in random Pauli basis in each layer. The final state is trashed, but the circuit realization $\scC$ (the gate choices and measurement observables) and the measurement outcomes $\vect{b}$ are recorded as a classical shadow. Repeated randomized measurements of copies of $\rho$ will collect a dataset of classical shadows, which can be used to predict the physical properties of the state $\rho$ through classical post-processing.}
\label{fig: protocol}
\end{center}
\end{figure}

Classical shadow tomography \cite{Huang2020P2002.08953, Ohliger2013E1204.5735, Guta2018F1809.11162} offers an efficient randomized measurement scheme for extracting physically relevant information from a quantum state. Much research \cite{Huang2021E2103.07510, Hadfield2020M2006.15788, Elben2020M2007.06305, Enshan-Koh2022C2011.11580, Hu2022H2102.10132, Hu2023C2107.04817, Levy2021C2110.02965, Bu2022C2202.03272, Hu2022L2203.07263, Seif2022S2203.07309, Hao-Low2022C2208.08964, Akhtar2023S2209.02093, Bertoni2022S2209.12924, Arienzo2022C2211.09835, Ippoliti2023O2212.11963} primarily concentrates on the randomized measurement protocol that entails random unitary evolution, followed by the final stage of local measurements on all qubits. This process is akin to halting the universe's time evolution to measure every qubit. A more realistic measurement scheme involves conducting local measurements intermittently while the entire quantum system continues to evolve, which more closely imitates how we observe the quantum universe surrounding us. This situation can be represented by hybrid quantum circuits \cite{Li2018Q1808.06134, Skinner2019M1808.05953, Potter2021E2111.08018, Fisher2023R2207.14280} formed by randomly interspersing local measurements among unitary gates in a quantum circuit. Notably, hybrid quantum circuits reveal a phase transition \cite{Li2019M1901.08092, Choi2020Q1903.05124, Gullans2020D1905.05195, Bao2020T1908.04305, Jian2020M1908.08051, Zabalo2020C1911.00008, Fan2021S2002.12385, Nahum2020M2009.11311,  Bao2021S2102.09164, Weinstein2022S2210.14242} in the quantum entanglement among qubits when the measurement rate surpasses a critical threshold, known as the measurement-induced entanglement transition or the purification transition. Our focus in this work is to explore the hybrid circuit as a randomized measurement scheme for classical shadow tomography and investigate the reconstruction of the quantum state using measurement outcomes obtained from intermittent measurements during the hybrid circuit's evolution, as illustrated in \figref{fig: protocol}.

The primary scientific question we aim to address concerns the efficiency of extracting information about the initial quantum state from intermittent measurement outcomes collected from the hybrid quantum circuit, within the context of classical shadow tomography. To address this problem, we first expanded the existing classical shadow tomography framework to accommodate more general scenarios where measurements can occur at any spacetime position throughout the quantum evolution. In particular, we introduced a systematic classical post-processing method for reconstructing the quantum state from the classical data of random circuit realizations and measurement outcomes in repeated experiments. Numerical simulations were conducted to validate the proposed reconstruction formula.

Subsequently, we defined the locally-scrambled shadow norm \cite{Hu2023C2107.04817, Bu2022C2202.03272} for the hybrid quantum circuit measurement scheme, which quantifies the typical number $M$ of experiments required to estimate the expectation value of an observable accurately, also referred to as the \emph{sample complexity} in quantum state tomography. Utilizing the tensor network method \cite{Akhtar2023S2209.02093, Bertoni2022S2209.12924}, we found that the sample complexity $M$ scales with the operator size $k$ of the observable as $M\simeq \beta^k\text{poly}(k)$, with the base $\beta$ depending on the measurement rate $p$ of the hybrid quantum circuit. We noted that $\beta$ is minimized (yielding optimal sample complexity scaling) when the measurement rate $p=p_c$ is tuned to the critical point of the measurement-induced transition in the hybrid quantum circuit. The minimal value is found to be around $\beta_\text{min}\approx 2.2$. Therefore, measurement-induced criticality is tomographically optimal within the scope of the hybrid quantum circuit measurement scheme.

\textit{Generalized Classical Snapshots.} --- The theoretical framework of classical shadow tomography can be extended to accommodate more general randomized measurement schemes \cite{Acharya2021I2105.05992, Chau-Nguyen2022O2205.08990} that permit intermittent and partial measurements throughout random quantum evolutions. Conceptually, the idea is as follows: irrespective of how single-qubit measurements are arranged and implemented in a single-shot experiment, the experimental result must be a string of classical bits, denoted as $\vect{b}=(b_1,b_2,\cdots)$, which represents the measurement outcome $b_n\in\{0,1\}$ for the $n$th measurement in the process. Given an initial quantum state $\rho$ and a particular measurement circuit $\scC$ (specified by both the circuit structure and gate choices), the entire measurement protocol can be characterized by the conditional probability $p(\vect{b}|\rho,\scC)$. 

The linearity of quantum mechanics implies that there must exist a measurement operator $\sigma_{\vect{b}|\scC}$ associated with each possible string of measurement outcomes $\vect{b}$, such that:
\eq{\label{eq: def sigma_b}
p(\vect{b}|\rho,\scC)\propto\Tr(\sigma_{\vect{b}|\scC}\rho).
}
We will call the operator $\sigma_{\vect{b}|\scC}$ a \emph{classical snapshot}. In the conventional classical shadow tomography, where the randomized measurement is implemented by first applying a random unitary transformation $U$ to the initial state $\rho$ (as $\rho\to U\rho U^\dagger$) and then measuring every qubit separately in the $Z$-basis, the classical snapshot $\sigma_{\vect{b}|\scC}$ reduces to the standard form of $\sigma_{\vect{b}|\scC}=U^\dagger\ket{\vect{b}}\bra{\vect{b}}U$. Beyond this conventional setup, \eqnref{eq: def sigma_b} provides a more general definition of classical snapshots when the measurement protocol is more involved. The classical snapshot $\sigma_{\vect{b}|\scC}$ should be a Hermitian positive semi-definite operator to ensure the real positivity of the conditional probability $p(\vect{b}|\rho,\scC)$. Given this property, it is natural to normalize $\sigma_{\vect{b}|\scC}$ such that $\Tr\sigma_{\vect{b}|\scC}=1$ \footnote{See \appref{app: formalism} for more rigorous treatment of the normalization.}, and view $\sigma_{\vect{b}|\scC}$ as another density matrix, called the classical snapshot state.

\textit{Hybrid Quantum Circuit Measurement.} ---
The hybrid quantum circuit measurement scheme is depicted in \figref{fig: protocol}. Starting from an $N$-qubit unknown quantum state $\rho$ of interest, apply the measurement and unitary layers alternately, where:
\begin{itemize}
\item Measurement layer: For each qubit independently, with probability $p$, choose to measure it in one of the three Pauli bases randomly. In the $l$-th measurement layer, suppose $A_l$ is the subset of qubits chosen to be measured. For each chosen qubit $i \in A_l$, let $P_i^{(l)} \in \{X_i, Y_i, Z_i\}$ be the choice of Pauli observable and $b_i^{(l)} \in \{0, 1\}$ be the corresponding measurement outcome. The measurement layer is described by the Kraus operator
\eq{
K_l^M = \prod_{i\in A_l}\frac{\id+(-)^{b_i^{(l)}}P_i^{(l)}}{2}.
}
\item Unitary layer: For every other nearest-two-qubit bond independently, apply a Clifford gate \cite{Gottesman1997S, Gottesman1998Tquant-ph/9807006} uniformly drawn from the two-qubit Clifford group. The Kraus operator for the $l$-th unitary layer is
\eq{
K_l^U=\left\{
\begin{array}{ll}
\prod_i U_{2i,2i+1}^{(l)} & l\in\text{even},\\
\prod_i U_{2i-1,2i}^{(l)} & l\in\text{odd},
\end{array}
\right.
}
which alternates between even and odd bonds with the layer index $l$ (such that the unitary gates form a brick-wall pattern as shown in \figref{fig: protocol}).
\end{itemize}
Packing the choice of measurement subsets $A_l$, Pauli observables $\vect{P}^{(l)}$ and Clifford gates $\vect{U}^{(l)}$ (for $l=1,2,\cdots$) altogether into the specification of a measurement circuit $\scC$, and gathering all the measurement outcomes $\vect{b}=\{\vect{b}^{(l)}\}$ together as a classical bit-string, the probability to observe $\vect{b}$ given $\scC$ is
\eq{
p(\vect{b}|\rho,\scC)=\Tr(K_{\vect{b}|\scC}\,\rho \,K_{\vect{b}|\scC}^\dagger),}
where $K_{\vect{b}|\scC}=\prod_{l}K_l^U K_l^M$ is the overall Krause operator.

\begin{figure}[htbp]
\begin{center}
\includegraphics[scale=0.65]{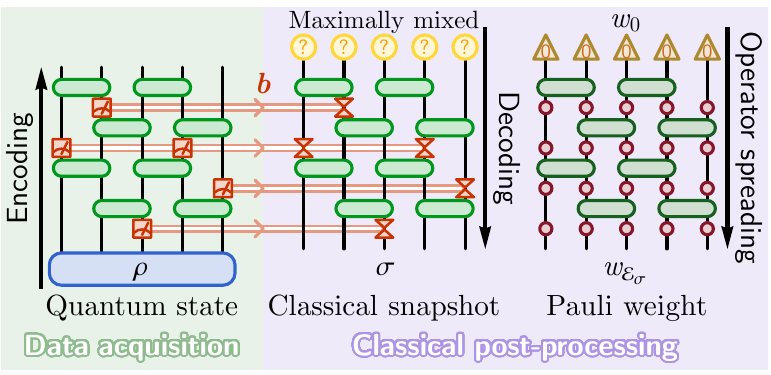}
\caption{Protocol of classical shadow tomography for hybrid quantum circuits. The quantum state $\rho$ is efficiently encoded as classical information by randomized measurements in the data acquisition phase. A classical snapshot state $\sigma$  is decoded by backward evolution from a maximally mixed state, given the circuit structure and measurement outcomes $\vect{b}$. On the other hand, its prior Pauli weights $w_{\scE_\sigma}(P)$ are inferred following the operator spreading dynamics. }
\label{fig: processing}
\end{center}
\end{figure}

Then following the assertion in \eqnref{eq: def sigma_b}, the classical snapshot associated with such measurement outcome should be identified as
\eq{\label{eq: def sigma}
\sigma_{\vect{b}|\scC}=\frac{K_{\vect{b}|\scC}^\dagger K_{\vect{b}|\scC}}{\Tr(K_{\vect{b}|\scC}^\dagger K_{\vect{b}|\scC}),}
}
where the denominator normalizes the classical snapshot as a state. Since the measurement circuit $\scC$ is composed of Clifford gates and Pauli measurements, every classical snapshot $\sigma_{\vect{b}|\scC}$ is a stabilizer state and can be efficiently represented and reconstructed on classical computers \cite{Gottesman1997S, Gottesman1998Tquant-ph/9807006}. As illustrated in \figref{fig: processing}, to reconstruct $\sigma_{\vect{b}|\scC}$, one starts with a maximally mixed state (described by the density matrix $\id/(\Tr \id)$) and traces back the measurement circuit: inverting every unitary gate, replacing every measurement by projection to the measurement outcome, and normalizing the final state in the end.

\textit{Posterior and Prior Distributions.} --- We can interpret the hybrid quantum circuit measurement process as a measure-and-prepare quantum channel that measures the initial state $\rho$ and prepares the classical snapshot state $\sigma$ with the \emph{posterior} probability:
\begin{equation}
p(\sigma|\rho) :=\sum_{\vect{b},\scC} \delta_{\sigma, \sigma_{\vect{b}|\scC}} p(\vect{b}|\rho,\scC)p(\scC),
\end{equation}
where $p(\scC)$ denotes the probability of realizing a specific measurement circuit $\scC$. Assuming that all Pauli measurements and Clifford gates are chosen uniformly, $p(\scC)\propto \prod_{l} p^{|A_l|}(1-p)^{N-|A_l|}$ will only be affected by the measurement rate $p$ of the hybrid quantum circuit. By conducting the hybrid quantum circuit measurements of the target state $\rho$ repeatedly, one can sample classical snapshot states $\sigma$ from the \emph{posterior} distribution $p(\sigma|\rho)$, forming an ensemble $\scE_{\sigma|\rho}=\{\sigma|\sigma\sim p(\sigma|\rho)\}$. The objective of classical shadow tomography is to predict properties of $\rho$ based on the samples of $\scE_{\sigma|\rho}$ collected from experiments as classical data. 

We introduce the \emph{prior} distribution $p(\sigma)$ of the classical snapshot \cite{Akhtar2023S2209.02093}, defined as $p(\sigma):=p(\sigma|\rho=\id/(\Tr\id))$. This distribution describes our knowledge about classical snapshots before observing the quantum state $\rho$ (as if $\rho$ is maximally mixed). The prior distribution solely characterizes the statistical properties of the randomized measurement scheme, reflecting our uncertainty about the measurement circuit structures and gate choices. 

\textit{Pauli weight.} --- A crucial property of the prior classical snapshot ensemble $\scE_\sigma=\{\sigma|\sigma\sim p(\sigma)\}$ is its \emph{Pauli weight} \cite{Bu2022C2202.03272, Bertoni2022S2209.12924}
\begin{equation}
w_{\scE_\sigma}(P):=\E_{\sigma\sim p(\sigma)}(\Tr P\sigma)^2,
\end{equation}
defined for any Pauli operator $P=\prod_i P_i$ (where $P_i\in\{I,X,Y,Z\}$ denotes the Pauli operator on the $i$-th qubit). The Pauli weight $w_{\scE_\sigma}(P)$ fully characterizes the second-moment statistical feature of the prior distribution $p(\sigma)$. It represents the probability for a Pauli observable $P$ to be transformed to the measurement basis and observed directly by the randomized measurement. It plays an important role in performing and analyzing classical shadow tomography.

For hybrid quantum circuits, the Pauli weight can be computed following the operator dynamics \cite{Ho2017E1508.03784, Bohrdt2017S1612.02434, Nahum2017Q1608.06950, Kukuljan2017W1701.09147, von-Keyserlingk2018O1705.08910, Nahum2018O1705.08975, Rakovszky2018D1710.09827, Khemani2018O1710.09835, Chan2018S1712.06836, Zhou2019O1805.09307, Zhou2019E1804.09737, Xu2019L1805.05376, Chen2019Q1808.09812, Parker2019A1812.08657, Qi2019M1906.00524, Kuo2020M1910.11351, Akhtar2020M2006.08797}. For every step of the physical evolution of a random quantum state $\rho$ through a random quantum channel $\scK$, the Pauli weight will be updated by the Markov process \footnote{See \appref{app: formalism} for a brief review of the Markov evolution of Pauli weights.}
\eq{\label{eq: op dyn}
w_{\scE_{\scK(\rho)}}(P)=\sum_{P'}w_{\scE_\scK}(P,P')w_{\scE_\rho}(P'),
}
where $w_{\scE_\scK}(P,P'):=\E_{\scK\in\scE_\scK}(\Tr(P\scK(P'))/\Tr\id)^2$ is the Pauli transfer matrix of the random channel ensemble $\scE_\scK$. For every two-qubit random Clifford unitary channel $\scU$ and every probabilistic single-qubit random Pauli measurement channel $\scM$, the corresponding Pauli transfer matrices are
\eqs{
w_{\scE_\scU}(P,P')&=\check{\delta}_{P,\id}\check{\delta}_{P',\id}+\frac{1}{15}(1-\check{\delta}_{P,\id})(1-\check{\delta}_{P',\id}),\\
w_{\scE_\scM}(P,P')&=\frac{p}{9}(1+2\check{\delta}_{P,\id})(1+2\check{\delta}_{P',\id})+(1-p)\check{\delta}_{P,P'},
}
where $\check{\delta}$ denotes the Kronecker delta symbol restricted to the support of the corresponding quantum channel. Starting from the initial Pauli weight $w_0(P)=\delta_{P,\id}$ of the maximally mixed state and applying the Pauli transfer matrix in accordance with the measurement circuit structure (see \figref{fig: processing}), the classical snapshot Pauli weight $w_{\scE_\sigma}(P)$ can be evaluated following \eqnref{eq: op dyn} \footnote{Strictly speaking, the classical snapshot Pauli weight dynamics is not Markovian due to the normalization factor on the denominator of \eqnref{eq: def sigma}. We are taking a Markovian approximation, that enables us to make thermodynamic-limit estimation of the shadow norm. The approximation is expected to work away from the measurement-induced critical point when the critical fluctuation is suppressed.}. In the end, the Pauli weight should be normalized to $w_{\scE_\sigma}(\id)=1$ to be consistent with the normalization of the classical snapshot states defined in \eqnref{eq: def sigma}.

\textit{Observable Estimation.} --- We now present a key result of our study: given any Pauli observable $P$, its expectation value on the initial quantum state $\rho$ can be inferred from the posterior classical snapshots via \cite{Bu2022C2202.03272, Bertoni2022S2209.12924}
\eq{\label{eq: P exp}
\langle P\rangle:=\Tr(P\rho)=\E_{\sigma\sim p(\sigma|\rho)}\frac{\Tr(P\sigma)}{w_{\scE_\sigma}(P)}.
}
For more general observable $O=\sum_P o_P P$, the expectation value can be similarly predicted by $\langle O\rangle=\E_{\sigma\sim p(\sigma|\rho)}O_\sigma$, where $O_\sigma:=\sum_P o_P\Tr(P\sigma)/w_{\scE_\sigma}(P)$ is the \emph{single-shot estimation} \cite{Huang2020P2002.08953} of the observable $O$ given a particular classical snapshot $\sigma$, defined  based on \eqnref{eq: P exp}. This allows us to decode the quantum information about the original state $\rho$ from the classical shadows collected from the hybrid quantum circuit measurement. In practice, the expectation $\E_{\sigma\sim p(\sigma|\rho)}$ is often estimated by the median of means over a finite number of classical snapshots collected from experiments. 

\begin{figure}[htbp]
\begin{center}
\includegraphics[scale=0.65]{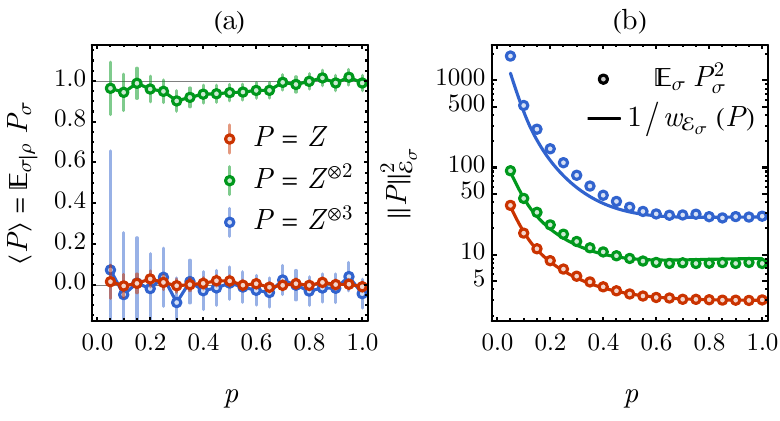}
\caption{Demonstration of hybrid quantum circuit classical shadow tomography on a 12-qubit GHZ state. (a) Predicted observable expectation values $\langle P\rangle$ and (b) locally-scrambled shadow norm $\Vert P\Vert_{\scE_\sigma}^2$ as functions of the measurement rate $p$. Colors label different Pauli observables $P=Z^{\otimes k}$.}
\label{fig: demo}
\end{center}
\end{figure}

To demonstrate the validity of \eqnref{eq: P exp}, we carried out numerical experiments. We take the Greenberger-Horne-Zeilinger (GHZ) \cite{Greenberger2007G0712.0921} state of $N=12$ qubits, described by $\rho=\ket{\psi}\bra{\psi}$ with $\ket{\psi}=(\ket{0}^{\otimes N}+\ket{1}^{\otimes N})/\sqrt{2}$. We consider a randomized measurement scheme implemented by shallow hybrid circuits, which contain three layers of random Clifford gates, together with random Pauli measurements inserted before each unitary layer with probability $p$ on each qubit. We simulate the protocol numerically on a classical computer by repeatedly preparing the GHZ state, applying the hybrid circuit, and collecting the measurement outcomes. For every given measurement rate $p$, we collect $M=50000$ samples and estimate the Pauli observables $P=Z^{\otimes k}$ based on the measurement outcomes using \eqnref{eq: P exp}. Our results, shown in \figref{fig: demo}(a), demonstrate that the estimated observable expectation values are consistent with their theoretical expectation on the GHZ state throughout the full range of $p$, i.e., $\langle Z^{\otimes k}\rangle=\frac{1}{2}(1+(-1)^k)$.

\textit{Sample Complexity Scaling.} --- The statistical uncertainty in the estimation, indicated by the error bar in \figref{fig: demo}(a), is due to the finite number of samples. The typical variance $\var \langle O\rangle \sim \Vert O\Vert_{\scE_\sigma}^2/M$ scales inversely with the number $M$ of samples. The coefficient $\Vert O\Vert_{\scE_\sigma}^2:=\E_{\sigma\sim p(\sigma)} O_\sigma^2$ is the \emph{locally-scrambled shadow norm}, introduced in \refcite{Hu2023C2107.04817}. It upper-bounds the variance of the single-shot estimation $O_\sigma$ over the prior classical snapshot ensemble $\scE_\sigma$. For Pauli observable $P$, the shadow norm has a simple expression \cite{Bu2022C2202.03272, Bertoni2022S2209.12924}
\eq{
\Vert P\Vert_{\scE_\sigma}^2=\frac{1}{w_{\scE_\sigma}(P)}.
}
In \figref{fig: demo}(b), the second moment of the single-shot estimation $\E_{\sigma\sim p(\sigma)} P_\sigma^2$ is compared with the inverse Pauli weight $1/w_{\scE_\sigma}(P)$ calculated from operator spreading dynamics. The results indicate a close match between the two measures. For generic observable $O=\sum_P o_P P$, the shadow norm is given by $\Vert O\Vert_{\scE_\sigma}^2=\sum_{P}|o_P|^2\Vert P\Vert_{\scE_\sigma}^2$. The shadow norm quantifies the number $M$ of samples needed to control the estimation variances $\var \langle{O}\rangle \lesssim\delta^2$ below a desired level set by a small $\delta$, which scales as $M\sim \Vert O\Vert_{\scE_\sigma}^2/\delta^2$. Therefore, the shadow norm measures the {sample complexity} for classical shadow tomography to predict the observable $O$ based on the randomized measurement scheme characterized by $\scE_\sigma$. 

\begin{figure}[htbp]
\begin{center}
\includegraphics[scale=0.65]{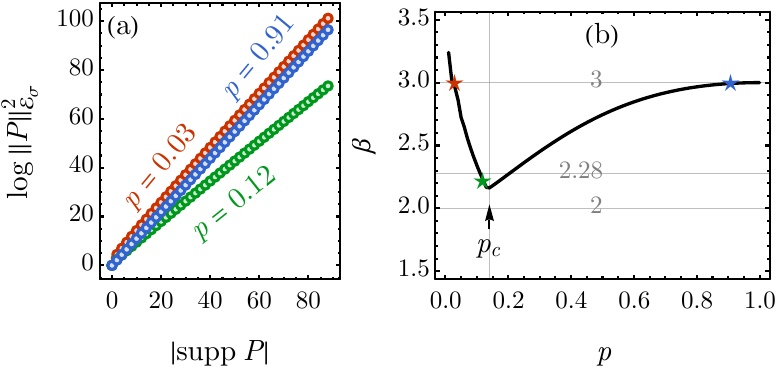}
\caption{(a) Dependence of log shadow norm $\log\Vert P\Vert_{\scE_\sigma}^2$ of consecutive Pauli string observable $P$ of size $k$ at different measurement rates $p$, demonstrating a leading linear behavior. (b) The base $\beta$ minimizes at a measurement rate $p_c$ that matches the measurement-induced transition of hybrid circuits. The measurement rates exemplified in (a) are highlighted as stars in (b).}
\label{fig: scaling}
\end{center}
\end{figure}

To study how the shadow norm scales with the operator size, we use the matrix product state (MPS) based approach developed in \refcite{Akhtar2023S2209.02093, Bertoni2022S2209.12924} to compute the Pauli weight $w_{\scE_\sigma}(P)$ following the operator spreading dynamics and determine the shadow norm $\Vert P\Vert_{\scE_\sigma}^2=1/w_{\scE_\sigma}(P)$ for \emph{consecutive} Pauli string observables $P$ of different sizes $k=|\supp P|$. The result is plotted in \figref{fig: scaling}(a). The shadow norm scales with the operator size $k$ exponentially with a base $\beta$ at the leading level
\eq{\Vert P\Vert_{\scE_\sigma}^2\simeq \beta^k\text{poly}(k),}
where $\text{poly}(k)$ stands for sub-leading correction that is polynomial in $k$. This is consistent with the intuition that longer Pauli observable will require exponentially more local measurements to determine. However, the base $\beta$ depends on the measurement rate $p$, as shown in \figref{fig: scaling}(b). 

We find that $\beta$ is minimized at $p=p_c$ when the hybrid quantum circuit operates at the measurement-induced criticality, and the shadow norm scales as 
\eq{\Vert P\Vert_{\scE_\sigma}^2|_{p=p_c}\simeq \beta_\text{min}^k k^{2\Delta},}
where $\beta_\text{min}=2.23\pm0.006$ and $\Delta=0.33\pm0.02$ are determined by fitting. We expect the critical exponent $\Delta$ to be universal, corresponding to the scaling dimension of a defect operator in the boundary conformal field theory (CFT) for the measurement-induced transition \footnote{See \appref{app: statmech} for a statistical mechanical model interpretation for the scaling behavior.}. The minimal $\beta_\text{min}$ enters the region between $3^{3/4}\approx 2.28$ and $2$, which is the range of optimal scaling achievable by shallow circuit classical shadows \cite{Ippoliti2023O2212.11963}. 

The minimization of $\beta$ can be understood by examining it from both sides of the phase transition. In the area-law phase ($p>p_c$), $\beta$ should decrease with decreasing measurement rate $p$. This is because a lower measurement rate allows for a few more local measurements to be deferred to deeper layers of the unitary circuit, enabling larger-size observables to be probed more efficiently by leveraging the scrambling power of shallow circuits. However, in the volume-law phase ($p<p_c$), if the measurement rate continues to decrease, $\beta$ will instead increase. Because the circuit's scrambling power becomes so strong that it begins to hide the quantum information of the initial state from local measurements deep in the circuit \cite{Choi2020Q1903.05124, Gullans2020D1905.05195, Fan2021S2002.12385}, which renders the measurements increasingly inefficient. As the measurement rate approaches zero ($p\to 0$), the shadow norm must diverge, because it becomes impossible to reconstruct the initial state in the absence of measurements. Therefore, the optimal scaling of the shadow norm (or the sample complexity) can only occur at the transition point $p_c$, where observables of all scales are probed efficiently \footnote{See \appref{app: toy} for a more detailed quantitative analysis using toy models.}.

\textit{Summary and Discussions.} --- In this work, we present the classical shadow tomography approach for decoding quantum information from measurement outcomes of hybrid quantum circuits. This method involves computing classical snapshots associated with measurement outcomes and using them to infer properties of the initial quantum state. The Pauli weight of the prior classical snapshot ensemble characterizes the statistical properties of the randomized measurement scheme, and the shadow norm quantifies the sample complexity for predicting observables. The log shadow norm scales linearly with the operator size of the observable and exhibits optimal scaling at a critical measurement rate of the hybrid circuit that corresponds to the measurement-induced criticality. 

Hybrid quantum circuits are known for their error correction encoding in the volume-law phase \cite{Choi2020Q1903.05124, Gullans2020D1905.05195, Fan2021S2002.12385}. To use them as a random quantum error correction code, the ability to decode quantum information from measurement outcomes is essential. Classical shadow tomography provides a systematic and universal way to decode hybrid quantum circuits, making them suitable for more exciting quantum information applications.

Measurement-induced transition in hybrid quantum circuits was originally proposed as an entanglement transition. However, measuring entanglement entropy is a difficult task that requires post-selections. With classical shadow tomography, we can directly benchmark the prior classical snapshot Pauli weight $w_{\scE_\sigma}(P)$ on a known quantum states $\rho$ (assuming $\Tr(P\rho)\neq 0$),
\eq{w_{\scE_\sigma}(P)=\frac{\E_{\sigma\sim p(\sigma|\rho)}\Tr(P\sigma)}{\Tr(P\rho)}.}
where $p(\sigma|\rho)$ can be sampled by performing the hybrid circuit measurement on $\rho$. Then $\beta$ can be extracted by fitting the dependence of $w_{\scE_\sigma}(P)$ with respect to its operator size $k$. It is supposed to exhibit a kink at the measurement-induced transition as \figref{fig: scaling}, which provides another method to detect the transition without post-selections apart from the cross-entropy benchmark \cite{Li2023C2209.00609}.

\textit{Note added.} --- Up on finishing this work, we become aware that a related work \cite{Ippoliti2023L2307.15011} appeared.

\begin{acknowledgements}
We acknowledge the helpful discussions with Ehud Altman, Matthew Fisher, Michael Gullans, Yaodong Li, and Bryan Clark.  We are especially grateful to Ehud Altman for inspiring us on the quantum statistical mechanical model understanding of our results. A.A.A. and Y.Z.Y. are supported by a startup fund from UCSD. H.Y.H. is grateful for the support by Harvard Quantum Initiative Fellowship.
\end{acknowledgements}

\bibliographystyle{apsrev4-1} 
\bibliography{ref}

\begin{thebibliography}{61}%
\makeatletter
\providecommand \@ifxundefined [1]{%
 \@ifx{#1\undefined}
}%
\providecommand \@ifnum [1]{%
 \ifnum #1\expandafter \@firstoftwo
 \else \expandafter \@secondoftwo
 \fi
}%
\providecommand \@ifx [1]{%
 \ifx #1\expandafter \@firstoftwo
 \else \expandafter \@secondoftwo
 \fi
}%
\providecommand \natexlab [1]{#1}%
\providecommand \enquote  [1]{``#1''}%
\providecommand \bibnamefont  [1]{#1}%
\providecommand \bibfnamefont [1]{#1}%
\providecommand \citenamefont [1]{#1}%
\providecommand \href@noop [0]{\@secondoftwo}%
\providecommand \href [0]{\begingroup \@sanitize@url \@href}%
\providecommand \@href[1]{\@@startlink{#1}\@@href}%
\providecommand \@@href[1]{\endgroup#1\@@endlink}%
\providecommand \@sanitize@url [0]{\catcode `\\12\catcode `\$12\catcode
  `\&12\catcode `\#12\catcode `\^12\catcode `\_12\catcode `\%12\relax}%
\providecommand \@@startlink[1]{}%
\providecommand \@@endlink[0]{}%
\providecommand \url  [0]{\begingroup\@sanitize@url \@url }%
\providecommand \@url [1]{\endgroup\@href {#1}{\urlprefix }}%
\providecommand \urlprefix  [0]{URL }%
\providecommand \Eprint [0]{\href }%
\providecommand \doibase [0]{http://dx.doi.org/}%
\providecommand \selectlanguage [0]{\@gobble}%
\providecommand \bibinfo  [0]{\@secondoftwo}%
\providecommand \bibfield  [0]{\@secondoftwo}%
\providecommand \translation [1]{[#1]}%
\providecommand \BibitemOpen [0]{}%
\providecommand \bibitemStop [0]{}%
\providecommand \bibitemNoStop [0]{.\EOS\space}%
\providecommand \EOS [0]{\spacefactor3000\relax}%
\providecommand \BibitemShut  [1]{\csname bibitem#1\endcsname}%
\let\auto@bib@innerbib\@empty
\bibitem [{\citenamefont {{Huang}}\ \emph {et~al.}(2020)\citenamefont
  {{Huang}}, \citenamefont {{Kueng}},\ and\ \citenamefont
  {{Preskill}}}]{Huang2020P2002.08953}%
  \BibitemOpen
  \bibfield  {author} {\bibinfo {author} {\bibfnamefont {H.-Y.}\ \bibnamefont
  {{Huang}}}, \bibinfo {author} {\bibfnamefont {R.}~\bibnamefont {{Kueng}}}, \
  and\ \bibinfo {author} {\bibfnamefont {J.}~\bibnamefont {{Preskill}}},\
  }\href {\doibase 10.1038/s41567-020-0932-7} {\bibfield  {journal} {\bibinfo
  {journal} {Nature Physics}\ }\textbf {\bibinfo {volume} {16}},\ \bibinfo
  {pages} {1050} (\bibinfo {year} {2020})},\ \Eprint
  {http://arxiv.org/abs/2002.08953} {arXiv:2002.08953 [quant-ph]} \BibitemShut
  {NoStop}%
\bibitem [{\citenamefont {{Ohliger}}\ \emph {et~al.}(2013)\citenamefont
  {{Ohliger}}, \citenamefont {{Nesme}},\ and\ \citenamefont
  {{Eisert}}}]{Ohliger2013E1204.5735}%
  \BibitemOpen
  \bibfield  {author} {\bibinfo {author} {\bibfnamefont {M.}~\bibnamefont
  {{Ohliger}}}, \bibinfo {author} {\bibfnamefont {V.}~\bibnamefont {{Nesme}}},
  \ and\ \bibinfo {author} {\bibfnamefont {J.}~\bibnamefont {{Eisert}}},\
  }\href {\doibase 10.1088/1367-2630/15/1/015024} {\bibfield  {journal}
  {\bibinfo  {journal} {New Journal of Physics}\ }\textbf {\bibinfo {volume}
  {15}},\ \bibinfo {eid} {015024} (\bibinfo {year} {2013})},\ \Eprint
  {http://arxiv.org/abs/1204.5735} {arXiv:1204.5735 [quant-ph]} \BibitemShut
  {NoStop}%
\bibitem [{\citenamefont {{Guta}}\ \emph {et~al.}(2020)\citenamefont {{Guta}},
  \citenamefont {{Kahn}}, \citenamefont {{Kueng}},\ and\ \citenamefont
  {{Tropp}}}]{Guta2018F1809.11162}%
  \BibitemOpen
  \bibfield  {author} {\bibinfo {author} {\bibfnamefont {M.}~\bibnamefont
  {{Guta}}}, \bibinfo {author} {\bibfnamefont {J.}~\bibnamefont {{Kahn}}},
  \bibinfo {author} {\bibfnamefont {R.}~\bibnamefont {{Kueng}}}, \ and\
  \bibinfo {author} {\bibfnamefont {J.~A.}\ \bibnamefont {{Tropp}}},\ }\href
  {\doibase 10.1088/1751-8121/ab8111} {\bibfield  {journal} {\bibinfo
  {journal} {Journal of Physics A: Mathematical and Theoretical}\ }\textbf
  {\bibinfo {volume} {53}},\ \bibinfo {pages} {204001} (\bibinfo {year}
  {2020})},\ \Eprint {http://arxiv.org/abs/1809.11162} {arXiv:1809.11162
  [quant-ph]} \BibitemShut {NoStop}%
\bibitem [{\citenamefont {{Huang}}\ \emph {et~al.}(2021)\citenamefont
  {{Huang}}, \citenamefont {{Kueng}},\ and\ \citenamefont
  {{Preskill}}}]{Huang2021E2103.07510}%
  \BibitemOpen
  \bibfield  {author} {\bibinfo {author} {\bibfnamefont {H.-Y.}\ \bibnamefont
  {{Huang}}}, \bibinfo {author} {\bibfnamefont {R.}~\bibnamefont {{Kueng}}}, \
  and\ \bibinfo {author} {\bibfnamefont {J.}~\bibnamefont {{Preskill}}},\
  }\href {\doibase 10.1103/PhysRevLett.127.030503} {\bibfield  {journal}
  {\bibinfo  {journal} {\prl}\ }\textbf {\bibinfo {volume} {127}},\ \bibinfo
  {eid} {030503} (\bibinfo {year} {2021})},\ \Eprint
  {http://arxiv.org/abs/2103.07510} {arXiv:2103.07510 [quant-ph]} \BibitemShut
  {NoStop}%
\bibitem [{\citenamefont {{Hadfield}}\ \emph {et~al.}(2022)\citenamefont
  {{Hadfield}}, \citenamefont {{Bravyi}}, \citenamefont {{Raymond}},\ and\
  \citenamefont {{Mezzacapo}}}]{Hadfield2020M2006.15788}%
  \BibitemOpen
  \bibfield  {author} {\bibinfo {author} {\bibfnamefont {C.}~\bibnamefont
  {{Hadfield}}}, \bibinfo {author} {\bibfnamefont {S.}~\bibnamefont
  {{Bravyi}}}, \bibinfo {author} {\bibfnamefont {R.}~\bibnamefont {{Raymond}}},
  \ and\ \bibinfo {author} {\bibfnamefont {A.}~\bibnamefont {{Mezzacapo}}},\
  }\href {https://doi.org/10.1007/s00220-022-04343-8} {\bibfield  {journal}
  {\bibinfo  {journal} {Communications in Mathematical Physics}\ }\textbf
  {\bibinfo {volume} {391}},\ \bibinfo {eid} {arXiv:2006.15788} (\bibinfo
  {year} {2022})},\ \Eprint {http://arxiv.org/abs/2006.15788} {arXiv:2006.15788
  [quant-ph]} \BibitemShut {NoStop}%
\bibitem [{\citenamefont {{Elben}}\ \emph {et~al.}(2020)\citenamefont
  {{Elben}}, \citenamefont {{Kueng}}, \citenamefont {{Huang}}, \citenamefont
  {{van Bijnen}}, \citenamefont {{Kokail}}, \citenamefont {{Dalmonte}},
  \citenamefont {{Calabrese}}, \citenamefont {{Kraus}}, \citenamefont
  {{Preskill}}, \citenamefont {{Zoller}},\ and\ \citenamefont
  {{Vermersch}}}]{Elben2020M2007.06305}%
  \BibitemOpen
  \bibfield  {author} {\bibinfo {author} {\bibfnamefont {A.}~\bibnamefont
  {{Elben}}}, \bibinfo {author} {\bibfnamefont {R.}~\bibnamefont {{Kueng}}},
  \bibinfo {author} {\bibfnamefont {H.-Y.~R.}\ \bibnamefont {{Huang}}},
  \bibinfo {author} {\bibfnamefont {R.}~\bibnamefont {{van Bijnen}}}, \bibinfo
  {author} {\bibfnamefont {C.}~\bibnamefont {{Kokail}}}, \bibinfo {author}
  {\bibfnamefont {M.}~\bibnamefont {{Dalmonte}}}, \bibinfo {author}
  {\bibfnamefont {P.}~\bibnamefont {{Calabrese}}}, \bibinfo {author}
  {\bibfnamefont {B.}~\bibnamefont {{Kraus}}}, \bibinfo {author} {\bibfnamefont
  {J.}~\bibnamefont {{Preskill}}}, \bibinfo {author} {\bibfnamefont
  {P.}~\bibnamefont {{Zoller}}}, \ and\ \bibinfo {author} {\bibfnamefont
  {B.}~\bibnamefont {{Vermersch}}},\ }\href {\doibase
  10.1103/PhysRevLett.125.200501} {\bibfield  {journal} {\bibinfo  {journal}
  {\prl}\ }\textbf {\bibinfo {volume} {125}},\ \bibinfo {eid} {200501}
  (\bibinfo {year} {2020})},\ \Eprint {http://arxiv.org/abs/2007.06305}
  {arXiv:2007.06305 [quant-ph]} \BibitemShut {NoStop}%
\bibitem [{\citenamefont {{Enshan Koh}}\ and\ \citenamefont
  {{Grewal}}(2022)}]{Enshan-Koh2022C2011.11580}%
  \BibitemOpen
  \bibfield  {author} {\bibinfo {author} {\bibfnamefont {D.}~\bibnamefont
  {{Enshan Koh}}}\ and\ \bibinfo {author} {\bibfnamefont {S.}~\bibnamefont
  {{Grewal}}},\ }\href {\doibase 10.48550/arXiv.2011.11580} {\bibfield
  {journal} {\bibinfo  {journal} {{Quantum}}\ }\textbf {\bibinfo {volume}
  {6}},\ \bibinfo {eid} {arXiv:2011.11580} (\bibinfo {year} {2022})},\ \Eprint
  {http://arxiv.org/abs/2011.11580} {arXiv:2011.11580 [quant-ph]} \BibitemShut
  {NoStop}%
\bibitem [{\citenamefont {{Hu}}\ and\ \citenamefont
  {{You}}(2022)}]{Hu2022H2102.10132}%
  \BibitemOpen
  \bibfield  {author} {\bibinfo {author} {\bibfnamefont {H.-Y.}\ \bibnamefont
  {{Hu}}}\ and\ \bibinfo {author} {\bibfnamefont {Y.-Z.}\ \bibnamefont
  {{You}}},\ }\href {\doibase 10.1103/PhysRevResearch.4.013054} {\bibfield
  {journal} {\bibinfo  {journal} {Physical Review Research}\ }\textbf {\bibinfo
  {volume} {4}},\ \bibinfo {eid} {013054} (\bibinfo {year} {2022})},\ \Eprint
  {http://arxiv.org/abs/2102.10132} {arXiv:2102.10132 [quant-ph]} \BibitemShut
  {NoStop}%
\bibitem [{\citenamefont {{Hu}}\ \emph {et~al.}(2023)\citenamefont {{Hu}},
  \citenamefont {{Choi}},\ and\ \citenamefont {{You}}}]{Hu2023C2107.04817}%
  \BibitemOpen
  \bibfield  {author} {\bibinfo {author} {\bibfnamefont {H.-Y.}\ \bibnamefont
  {{Hu}}}, \bibinfo {author} {\bibfnamefont {S.}~\bibnamefont {{Choi}}}, \ and\
  \bibinfo {author} {\bibfnamefont {Y.-Z.}\ \bibnamefont {{You}}},\ }\href
  {\doibase 10.1103/PhysRevResearch.5.023027} {\bibfield  {journal} {\bibinfo
  {journal} {Physical Review Research}\ }\textbf {\bibinfo {volume} {5}},\
  \bibinfo {eid} {arXiv:2107.04817} (\bibinfo {year} {2023})},\ \Eprint
  {http://arxiv.org/abs/2107.04817} {arXiv:2107.04817 [quant-ph]} \BibitemShut
  {NoStop}%
\bibitem [{\citenamefont {{Levy}}\ \emph {et~al.}(2021)\citenamefont {{Levy}},
  \citenamefont {{Luo}},\ and\ \citenamefont {{Clark}}}]{Levy2021C2110.02965}%
  \BibitemOpen
  \bibfield  {author} {\bibinfo {author} {\bibfnamefont {R.}~\bibnamefont
  {{Levy}}}, \bibinfo {author} {\bibfnamefont {D.}~\bibnamefont {{Luo}}}, \
  and\ \bibinfo {author} {\bibfnamefont {B.~K.}\ \bibnamefont {{Clark}}},\
  }\href@noop {} {\bibfield  {journal} {\bibinfo  {journal} {arXiv e-prints}\
  ,\ \bibinfo {eid} {arXiv:2110.02965}} (\bibinfo {year} {2021})},\ \Eprint
  {http://arxiv.org/abs/2110.02965} {arXiv:2110.02965 [quant-ph]} \BibitemShut
  {NoStop}%
\bibitem [{\citenamefont {{Bu}}\ \emph {et~al.}(2022)\citenamefont {{Bu}},
  \citenamefont {{Enshan Koh}}, \citenamefont {{Garcia}},\ and\ \citenamefont
  {{Jaffe}}}]{Bu2022C2202.03272}%
  \BibitemOpen
  \bibfield  {author} {\bibinfo {author} {\bibfnamefont {K.}~\bibnamefont
  {{Bu}}}, \bibinfo {author} {\bibfnamefont {D.}~\bibnamefont {{Enshan Koh}}},
  \bibinfo {author} {\bibfnamefont {R.~J.}\ \bibnamefont {{Garcia}}}, \ and\
  \bibinfo {author} {\bibfnamefont {A.}~\bibnamefont {{Jaffe}}},\ }\href
  {\doibase 10.48550/arXiv.2202.03272} {\bibfield  {journal} {\bibinfo
  {journal} {arXiv e-prints}\ ,\ \bibinfo {eid} {arXiv:2202.03272}} (\bibinfo
  {year} {2022})},\ \Eprint {http://arxiv.org/abs/2202.03272} {arXiv:2202.03272
  [quant-ph]} \BibitemShut {NoStop}%
\bibitem [{\citenamefont {{Hu}}\ \emph {et~al.}(2022)\citenamefont {{Hu}},
  \citenamefont {{LaRose}}, \citenamefont {{You}}, \citenamefont {{Rieffel}},\
  and\ \citenamefont {{Wang}}}]{Hu2022L2203.07263}%
  \BibitemOpen
  \bibfield  {author} {\bibinfo {author} {\bibfnamefont {H.-Y.}\ \bibnamefont
  {{Hu}}}, \bibinfo {author} {\bibfnamefont {R.}~\bibnamefont {{LaRose}}},
  \bibinfo {author} {\bibfnamefont {Y.-Z.}\ \bibnamefont {{You}}}, \bibinfo
  {author} {\bibfnamefont {E.}~\bibnamefont {{Rieffel}}}, \ and\ \bibinfo
  {author} {\bibfnamefont {Z.}~\bibnamefont {{Wang}}},\ }\href@noop {}
  {\bibfield  {journal} {\bibinfo  {journal} {arXiv e-prints}\ ,\ \bibinfo
  {eid} {arXiv:2203.07263}} (\bibinfo {year} {2022})},\ \Eprint
  {http://arxiv.org/abs/2203.07263} {arXiv:2203.07263 [quant-ph]} \BibitemShut
  {NoStop}%
\bibitem [{\citenamefont {{Seif}}\ \emph {et~al.}(2022)\citenamefont {{Seif}},
  \citenamefont {{Cian}}, \citenamefont {{Zhou}}, \citenamefont {{Chen}},\ and\
  \citenamefont {{Jiang}}}]{Seif2022S2203.07309}%
  \BibitemOpen
  \bibfield  {author} {\bibinfo {author} {\bibfnamefont {A.}~\bibnamefont
  {{Seif}}}, \bibinfo {author} {\bibfnamefont {Z.-P.}\ \bibnamefont {{Cian}}},
  \bibinfo {author} {\bibfnamefont {S.}~\bibnamefont {{Zhou}}}, \bibinfo
  {author} {\bibfnamefont {S.}~\bibnamefont {{Chen}}}, \ and\ \bibinfo {author}
  {\bibfnamefont {L.}~\bibnamefont {{Jiang}}},\ }\href@noop {} {\bibfield
  {journal} {\bibinfo  {journal} {arXiv e-prints}\ ,\ \bibinfo {eid}
  {arXiv:2203.07309}} (\bibinfo {year} {2022})},\ \Eprint
  {http://arxiv.org/abs/2203.07309} {arXiv:2203.07309 [quant-ph]} \BibitemShut
  {NoStop}%
\bibitem [{\citenamefont {{Hao Low}}(2022)}]{Hao-Low2022C2208.08964}%
  \BibitemOpen
  \bibfield  {author} {\bibinfo {author} {\bibfnamefont {G.}~\bibnamefont {{Hao
  Low}}},\ }\href@noop {} {\bibfield  {journal} {\bibinfo  {journal} {arXiv
  e-prints}\ ,\ \bibinfo {eid} {arXiv:2208.08964}} (\bibinfo {year} {2022})},\
  \Eprint {http://arxiv.org/abs/2208.08964} {arXiv:2208.08964 [quant-ph]}
  \BibitemShut {NoStop}%
\bibitem [{\citenamefont {{Akhtar}}\ \emph {et~al.}(2023)\citenamefont
  {{Akhtar}}, \citenamefont {{Hu}},\ and\ \citenamefont
  {{You}}}]{Akhtar2023S2209.02093}%
  \BibitemOpen
  \bibfield  {author} {\bibinfo {author} {\bibfnamefont {A.~A.}\ \bibnamefont
  {{Akhtar}}}, \bibinfo {author} {\bibfnamefont {H.-Y.}\ \bibnamefont {{Hu}}},
  \ and\ \bibinfo {author} {\bibfnamefont {Y.-Z.}\ \bibnamefont {{You}}},\
  }\href {\doibase 10.22331/q-2023-06-01-1026} {\bibfield  {journal} {\bibinfo
  {journal} {Quantum}\ }\textbf {\bibinfo {volume} {7}},\ \bibinfo {pages}
  {1026} (\bibinfo {year} {2023})},\ \Eprint {http://arxiv.org/abs/2209.02093}
  {arXiv:2209.02093 [quant-ph]} \BibitemShut {NoStop}%
\bibitem [{\citenamefont {{Bertoni}}\ \emph {et~al.}(2022)\citenamefont
  {{Bertoni}}, \citenamefont {{Haferkamp}}, \citenamefont {{Hinsche}},
  \citenamefont {{Ioannou}}, \citenamefont {{Eisert}},\ and\ \citenamefont
  {{Pashayan}}}]{Bertoni2022S2209.12924}%
  \BibitemOpen
  \bibfield  {author} {\bibinfo {author} {\bibfnamefont {C.}~\bibnamefont
  {{Bertoni}}}, \bibinfo {author} {\bibfnamefont {J.}~\bibnamefont
  {{Haferkamp}}}, \bibinfo {author} {\bibfnamefont {M.}~\bibnamefont
  {{Hinsche}}}, \bibinfo {author} {\bibfnamefont {M.}~\bibnamefont
  {{Ioannou}}}, \bibinfo {author} {\bibfnamefont {J.}~\bibnamefont {{Eisert}}},
  \ and\ \bibinfo {author} {\bibfnamefont {H.}~\bibnamefont {{Pashayan}}},\
  }\href@noop {} {\bibfield  {journal} {\bibinfo  {journal} {arXiv e-prints}\
  ,\ \bibinfo {eid} {arXiv:2209.12924}} (\bibinfo {year} {2022})},\ \Eprint
  {http://arxiv.org/abs/2209.12924} {arXiv:2209.12924 [quant-ph]} \BibitemShut
  {NoStop}%
\bibitem [{\citenamefont {{Arienzo}}\ \emph {et~al.}(2022)\citenamefont
  {{Arienzo}}, \citenamefont {{Heinrich}}, \citenamefont {{Roth}},\ and\
  \citenamefont {{Kliesch}}}]{Arienzo2022C2211.09835}%
  \BibitemOpen
  \bibfield  {author} {\bibinfo {author} {\bibfnamefont {M.}~\bibnamefont
  {{Arienzo}}}, \bibinfo {author} {\bibfnamefont {M.}~\bibnamefont
  {{Heinrich}}}, \bibinfo {author} {\bibfnamefont {I.}~\bibnamefont {{Roth}}},
  \ and\ \bibinfo {author} {\bibfnamefont {M.}~\bibnamefont {{Kliesch}}},\
  }\href {\doibase 10.48550/arXiv.2211.09835} {\bibfield  {journal} {\bibinfo
  {journal} {arXiv e-prints}\ ,\ \bibinfo {eid} {arXiv:2211.09835}} (\bibinfo
  {year} {2022})},\ \Eprint {http://arxiv.org/abs/2211.09835} {arXiv:2211.09835
  [quant-ph]} \BibitemShut {NoStop}%
\bibitem [{\citenamefont {{Ippoliti}}\ \emph {et~al.}(2023)\citenamefont
  {{Ippoliti}}, \citenamefont {{Li}}, \citenamefont {{Rakovszky}},\ and\
  \citenamefont {{Khemani}}}]{Ippoliti2023O2212.11963}%
  \BibitemOpen
  \bibfield  {author} {\bibinfo {author} {\bibfnamefont {M.}~\bibnamefont
  {{Ippoliti}}}, \bibinfo {author} {\bibfnamefont {Y.}~\bibnamefont {{Li}}},
  \bibinfo {author} {\bibfnamefont {T.}~\bibnamefont {{Rakovszky}}}, \ and\
  \bibinfo {author} {\bibfnamefont {V.}~\bibnamefont {{Khemani}}},\ }\href
  {\doibase 10.1103/PhysRevLett.130.230403} {\bibfield  {journal} {\bibinfo
  {journal} {\prl}\ }\textbf {\bibinfo {volume} {130}},\ \bibinfo {eid}
  {230403} (\bibinfo {year} {2023})},\ \Eprint
  {http://arxiv.org/abs/2212.11963} {arXiv:2212.11963 [quant-ph]} \BibitemShut
  {NoStop}%
\bibitem [{\citenamefont {{Li}}\ \emph {et~al.}(2018)\citenamefont {{Li}},
  \citenamefont {{Chen}},\ and\ \citenamefont {{Fisher}}}]{Li2018Q1808.06134}%
  \BibitemOpen
  \bibfield  {author} {\bibinfo {author} {\bibfnamefont {Y.}~\bibnamefont
  {{Li}}}, \bibinfo {author} {\bibfnamefont {X.}~\bibnamefont {{Chen}}}, \ and\
  \bibinfo {author} {\bibfnamefont {M.~P.~A.}\ \bibnamefont {{Fisher}}},\
  }\href {\doibase 10.1103/PhysRevB.98.205136} {\bibfield  {journal} {\bibinfo
  {journal} {\prb}\ }\textbf {\bibinfo {volume} {98}},\ \bibinfo {eid} {205136}
  (\bibinfo {year} {2018})},\ \Eprint {http://arxiv.org/abs/1808.06134}
  {arXiv:1808.06134 [quant-ph]} \BibitemShut {NoStop}%
\bibitem [{\citenamefont {{Skinner}}\ \emph {et~al.}(2019)\citenamefont
  {{Skinner}}, \citenamefont {{Ruhman}},\ and\ \citenamefont
  {{Nahum}}}]{Skinner2019M1808.05953}%
  \BibitemOpen
  \bibfield  {author} {\bibinfo {author} {\bibfnamefont {B.}~\bibnamefont
  {{Skinner}}}, \bibinfo {author} {\bibfnamefont {J.}~\bibnamefont {{Ruhman}}},
  \ and\ \bibinfo {author} {\bibfnamefont {A.}~\bibnamefont {{Nahum}}},\ }\href
  {\doibase 10.1103/PhysRevX.9.031009} {\bibfield  {journal} {\bibinfo
  {journal} {Physical Review X}\ }\textbf {\bibinfo {volume} {9}},\ \bibinfo
  {eid} {031009} (\bibinfo {year} {2019})},\ \Eprint
  {http://arxiv.org/abs/1808.05953} {arXiv:1808.05953 [cond-mat.stat-mech]}
  \BibitemShut {NoStop}%
\bibitem [{\citenamefont {{Potter}}\ and\ \citenamefont
  {{Vasseur}}(2021)}]{Potter2021E2111.08018}%
  \BibitemOpen
  \bibfield  {author} {\bibinfo {author} {\bibfnamefont {A.~C.}\ \bibnamefont
  {{Potter}}}\ and\ \bibinfo {author} {\bibfnamefont {R.}~\bibnamefont
  {{Vasseur}}},\ }\href {\doibase 10.48550/arXiv.2111.08018} {\bibfield
  {journal} {\bibinfo  {journal} {arXiv e-prints}\ ,\ \bibinfo {eid}
  {arXiv:2111.08018}} (\bibinfo {year} {2021})},\ \Eprint
  {http://arxiv.org/abs/2111.08018} {arXiv:2111.08018 [quant-ph]} \BibitemShut
  {NoStop}%
\bibitem [{\citenamefont {{Fisher}}\ \emph {et~al.}(2023)\citenamefont
  {{Fisher}}, \citenamefont {{Khemani}}, \citenamefont {{Nahum}},\ and\
  \citenamefont {{Vijay}}}]{Fisher2023R2207.14280}%
  \BibitemOpen
  \bibfield  {author} {\bibinfo {author} {\bibfnamefont {M.~P.~A.}\
  \bibnamefont {{Fisher}}}, \bibinfo {author} {\bibfnamefont {V.}~\bibnamefont
  {{Khemani}}}, \bibinfo {author} {\bibfnamefont {A.}~\bibnamefont {{Nahum}}},
  \ and\ \bibinfo {author} {\bibfnamefont {S.}~\bibnamefont {{Vijay}}},\ }\href
  {\doibase 10.1146/annurev-conmatphys-031720-030658} {\bibfield  {journal}
  {\bibinfo  {journal} {Annual Review of Condensed Matter Physics}\ }\textbf
  {\bibinfo {volume} {14}},\ \bibinfo {pages} {335} (\bibinfo {year} {2023})},\
  \Eprint {http://arxiv.org/abs/2207.14280} {arXiv:2207.14280 [quant-ph]}
  \BibitemShut {NoStop}%
\bibitem [{\citenamefont {{Li}}\ \emph {et~al.}(2019)\citenamefont {{Li}},
  \citenamefont {{Chen}},\ and\ \citenamefont {{Fisher}}}]{Li2019M1901.08092}%
  \BibitemOpen
  \bibfield  {author} {\bibinfo {author} {\bibfnamefont {Y.}~\bibnamefont
  {{Li}}}, \bibinfo {author} {\bibfnamefont {X.}~\bibnamefont {{Chen}}}, \ and\
  \bibinfo {author} {\bibfnamefont {M.~P.~A.}\ \bibnamefont {{Fisher}}},\
  }\href {\doibase 10.1103/PhysRevB.100.134306} {\bibfield  {journal} {\bibinfo
   {journal} {\prb}\ }\textbf {\bibinfo {volume} {100}},\ \bibinfo {eid}
  {134306} (\bibinfo {year} {2019})},\ \Eprint
  {http://arxiv.org/abs/1901.08092} {arXiv:1901.08092 [cond-mat.stat-mech]}
  \BibitemShut {NoStop}%
\bibitem [{\citenamefont {{Choi}}\ \emph {et~al.}(2020)\citenamefont {{Choi}},
  \citenamefont {{Bao}}, \citenamefont {{Qi}},\ and\ \citenamefont
  {{Altman}}}]{Choi2020Q1903.05124}%
  \BibitemOpen
  \bibfield  {author} {\bibinfo {author} {\bibfnamefont {S.}~\bibnamefont
  {{Choi}}}, \bibinfo {author} {\bibfnamefont {Y.}~\bibnamefont {{Bao}}},
  \bibinfo {author} {\bibfnamefont {X.-L.}\ \bibnamefont {{Qi}}}, \ and\
  \bibinfo {author} {\bibfnamefont {E.}~\bibnamefont {{Altman}}},\ }\href
  {\doibase 10.1103/PhysRevLett.125.030505} {\bibfield  {journal} {\bibinfo
  {journal} {\prl}\ }\textbf {\bibinfo {volume} {125}},\ \bibinfo {eid}
  {030505} (\bibinfo {year} {2020})},\ \Eprint
  {http://arxiv.org/abs/1903.05124} {arXiv:1903.05124 [quant-ph]} \BibitemShut
  {NoStop}%
\bibitem [{\citenamefont {{Gullans}}\ and\ \citenamefont
  {{Huse}}(2020)}]{Gullans2020D1905.05195}%
  \BibitemOpen
  \bibfield  {author} {\bibinfo {author} {\bibfnamefont {M.~J.}\ \bibnamefont
  {{Gullans}}}\ and\ \bibinfo {author} {\bibfnamefont {D.~A.}\ \bibnamefont
  {{Huse}}},\ }\href {\doibase 10.1103/PhysRevX.10.041020} {\bibfield
  {journal} {\bibinfo  {journal} {Physical Review X}\ }\textbf {\bibinfo
  {volume} {10}},\ \bibinfo {eid} {041020} (\bibinfo {year} {2020})},\ \Eprint
  {http://arxiv.org/abs/1905.05195} {arXiv:1905.05195 [quant-ph]} \BibitemShut
  {NoStop}%
\bibitem [{\citenamefont {{Bao}}\ \emph {et~al.}(2020)\citenamefont {{Bao}},
  \citenamefont {{Choi}},\ and\ \citenamefont {{Altman}}}]{Bao2020T1908.04305}%
  \BibitemOpen
  \bibfield  {author} {\bibinfo {author} {\bibfnamefont {Y.}~\bibnamefont
  {{Bao}}}, \bibinfo {author} {\bibfnamefont {S.}~\bibnamefont {{Choi}}}, \
  and\ \bibinfo {author} {\bibfnamefont {E.}~\bibnamefont {{Altman}}},\ }\href
  {\doibase 10.1103/PhysRevB.101.104301} {\bibfield  {journal} {\bibinfo
  {journal} {\prb}\ }\textbf {\bibinfo {volume} {101}},\ \bibinfo {eid}
  {104301} (\bibinfo {year} {2020})},\ \Eprint
  {http://arxiv.org/abs/1908.04305} {arXiv:1908.04305 [cond-mat.stat-mech]}
  \BibitemShut {NoStop}%
\bibitem [{\citenamefont {{Jian}}\ \emph {et~al.}(2020)\citenamefont {{Jian}},
  \citenamefont {{You}}, \citenamefont {{Vasseur}},\ and\ \citenamefont
  {{Ludwig}}}]{Jian2020M1908.08051}%
  \BibitemOpen
  \bibfield  {author} {\bibinfo {author} {\bibfnamefont {C.-M.}\ \bibnamefont
  {{Jian}}}, \bibinfo {author} {\bibfnamefont {Y.-Z.}\ \bibnamefont {{You}}},
  \bibinfo {author} {\bibfnamefont {R.}~\bibnamefont {{Vasseur}}}, \ and\
  \bibinfo {author} {\bibfnamefont {A.~W.~W.}\ \bibnamefont {{Ludwig}}},\
  }\href {\doibase 10.1103/PhysRevB.101.104302} {\bibfield  {journal} {\bibinfo
   {journal} {\prb}\ }\textbf {\bibinfo {volume} {101}},\ \bibinfo {eid}
  {104302} (\bibinfo {year} {2020})},\ \Eprint
  {http://arxiv.org/abs/1908.08051} {arXiv:1908.08051 [cond-mat.stat-mech]}
  \BibitemShut {NoStop}%
\bibitem [{\citenamefont {{Zabalo}}\ \emph {et~al.}(2020)\citenamefont
  {{Zabalo}}, \citenamefont {{Gullans}}, \citenamefont {{Wilson}},
  \citenamefont {{Gopalakrishnan}}, \citenamefont {{Huse}},\ and\ \citenamefont
  {{Pixley}}}]{Zabalo2020C1911.00008}%
  \BibitemOpen
  \bibfield  {author} {\bibinfo {author} {\bibfnamefont {A.}~\bibnamefont
  {{Zabalo}}}, \bibinfo {author} {\bibfnamefont {M.~J.}\ \bibnamefont
  {{Gullans}}}, \bibinfo {author} {\bibfnamefont {J.~H.}\ \bibnamefont
  {{Wilson}}}, \bibinfo {author} {\bibfnamefont {S.}~\bibnamefont
  {{Gopalakrishnan}}}, \bibinfo {author} {\bibfnamefont {D.~A.}\ \bibnamefont
  {{Huse}}}, \ and\ \bibinfo {author} {\bibfnamefont {J.~H.}\ \bibnamefont
  {{Pixley}}},\ }\href {\doibase 10.1103/PhysRevB.101.060301} {\bibfield
  {journal} {\bibinfo  {journal} {\prb}\ }\textbf {\bibinfo {volume} {101}},\
  \bibinfo {eid} {060301} (\bibinfo {year} {2020})},\ \Eprint
  {http://arxiv.org/abs/1911.00008} {arXiv:1911.00008 [cond-mat.dis-nn]}
  \BibitemShut {NoStop}%
\bibitem [{\citenamefont {{Fan}}\ \emph {et~al.}(2021)\citenamefont {{Fan}},
  \citenamefont {{Vijay}}, \citenamefont {{Vishwanath}},\ and\ \citenamefont
  {{You}}}]{Fan2021S2002.12385}%
  \BibitemOpen
  \bibfield  {author} {\bibinfo {author} {\bibfnamefont {R.}~\bibnamefont
  {{Fan}}}, \bibinfo {author} {\bibfnamefont {S.}~\bibnamefont {{Vijay}}},
  \bibinfo {author} {\bibfnamefont {A.}~\bibnamefont {{Vishwanath}}}, \ and\
  \bibinfo {author} {\bibfnamefont {Y.-Z.}\ \bibnamefont {{You}}},\ }\href
  {\doibase 10.1103/PhysRevB.103.174309} {\bibfield  {journal} {\bibinfo
  {journal} {\prb}\ }\textbf {\bibinfo {volume} {103}},\ \bibinfo {eid}
  {174309} (\bibinfo {year} {2021})},\ \Eprint
  {http://arxiv.org/abs/2002.12385} {arXiv:2002.12385 [cond-mat.stat-mech]}
  \BibitemShut {NoStop}%
\bibitem [{\citenamefont {{Nahum}}\ \emph {et~al.}(2020)\citenamefont
  {{Nahum}}, \citenamefont {{Roy}}, \citenamefont {{Skinner}},\ and\
  \citenamefont {{Ruhman}}}]{Nahum2020M2009.11311}%
  \BibitemOpen
  \bibfield  {author} {\bibinfo {author} {\bibfnamefont {A.}~\bibnamefont
  {{Nahum}}}, \bibinfo {author} {\bibfnamefont {S.}~\bibnamefont {{Roy}}},
  \bibinfo {author} {\bibfnamefont {B.}~\bibnamefont {{Skinner}}}, \ and\
  \bibinfo {author} {\bibfnamefont {J.}~\bibnamefont {{Ruhman}}},\ }\href
  {\doibase 10.48550/arXiv.2009.11311} {\bibfield  {journal} {\bibinfo
  {journal} {arXiv e-prints}\ ,\ \bibinfo {eid} {arXiv:2009.11311}} (\bibinfo
  {year} {2020})},\ \Eprint {http://arxiv.org/abs/2009.11311} {arXiv:2009.11311
  [cond-mat.stat-mech]} \BibitemShut {NoStop}%
\bibitem [{\citenamefont {{Bao}}\ \emph {et~al.}(2021)\citenamefont {{Bao}},
  \citenamefont {{Choi}},\ and\ \citenamefont {{Altman}}}]{Bao2021S2102.09164}%
  \BibitemOpen
  \bibfield  {author} {\bibinfo {author} {\bibfnamefont {Y.}~\bibnamefont
  {{Bao}}}, \bibinfo {author} {\bibfnamefont {S.}~\bibnamefont {{Choi}}}, \
  and\ \bibinfo {author} {\bibfnamefont {E.}~\bibnamefont {{Altman}}},\ }\href
  {\doibase 10.1016/j.aop.2021.168618} {\bibfield  {journal} {\bibinfo
  {journal} {Annals of Physics}\ }\textbf {\bibinfo {volume} {435}},\ \bibinfo
  {eid} {168618} (\bibinfo {year} {2021})},\ \Eprint
  {http://arxiv.org/abs/2102.09164} {arXiv:2102.09164 [cond-mat.stat-mech]}
  \BibitemShut {NoStop}%
\bibitem [{\citenamefont {{Weinstein}}\ \emph {et~al.}(2022)\citenamefont
  {{Weinstein}}, \citenamefont {{Kelly}}, \citenamefont {{Marino}},\ and\
  \citenamefont {{Altman}}}]{Weinstein2022S2210.14242}%
  \BibitemOpen
  \bibfield  {author} {\bibinfo {author} {\bibfnamefont {Z.}~\bibnamefont
  {{Weinstein}}}, \bibinfo {author} {\bibfnamefont {S.~P.}\ \bibnamefont
  {{Kelly}}}, \bibinfo {author} {\bibfnamefont {J.}~\bibnamefont {{Marino}}}, \
  and\ \bibinfo {author} {\bibfnamefont {E.}~\bibnamefont {{Altman}}},\ }\href
  {\doibase 10.48550/arXiv.2210.14242} {\bibfield  {journal} {\bibinfo
  {journal} {arXiv e-prints}\ ,\ \bibinfo {eid} {arXiv:2210.14242}} (\bibinfo
  {year} {2022})},\ \Eprint {http://arxiv.org/abs/2210.14242} {arXiv:2210.14242
  [quant-ph]} \BibitemShut {NoStop}%
\bibitem [{\citenamefont {{Acharya}}\ \emph {et~al.}(2021)\citenamefont
  {{Acharya}}, \citenamefont {{Saha}},\ and\ \citenamefont
  {{Sengupta}}}]{Acharya2021I2105.05992}%
  \BibitemOpen
  \bibfield  {author} {\bibinfo {author} {\bibfnamefont {A.}~\bibnamefont
  {{Acharya}}}, \bibinfo {author} {\bibfnamefont {S.}~\bibnamefont {{Saha}}}, \
  and\ \bibinfo {author} {\bibfnamefont {A.~M.}\ \bibnamefont {{Sengupta}}},\
  }\href {\doibase 10.1103/PhysRevA.104.052418} {\bibfield  {journal} {\bibinfo
   {journal} {{Phys. Rev. A}}\ }\textbf {\bibinfo {volume} {{104}}},\ \bibinfo
  {pages} {{052418}} (\bibinfo {year} {2021})},\ \Eprint
  {http://arxiv.org/abs/2105.05992} {arXiv:2105.05992 [quant-ph]} \BibitemShut
  {NoStop}%
\bibitem [{\citenamefont {{Chau Nguyen}}\ \emph {et~al.}(2022)\citenamefont
  {{Chau Nguyen}}, \citenamefont {{Lennart B{\"o}nsel}}, \citenamefont
  {{Steinberg}},\ and\ \citenamefont
  {{G{\"u}hne}}}]{Chau-Nguyen2022O2205.08990}%
  \BibitemOpen
  \bibfield  {author} {\bibinfo {author} {\bibfnamefont {H.}~\bibnamefont
  {{Chau Nguyen}}}, \bibinfo {author} {\bibfnamefont {J.}~\bibnamefont
  {{Lennart B{\"o}nsel}}}, \bibinfo {author} {\bibfnamefont {J.}~\bibnamefont
  {{Steinberg}}}, \ and\ \bibinfo {author} {\bibfnamefont {O.}~\bibnamefont
  {{G{\"u}hne}}},\ }\href {\doibase 10.48550/arXiv.2205.08990} {\bibfield
  {journal} {\bibinfo  {journal} {arXiv e-prints}\ ,\ \bibinfo {eid}
  {arXiv:2205.08990}} (\bibinfo {year} {2022})},\ \Eprint
  {http://arxiv.org/abs/2205.08990} {arXiv:2205.08990 [quant-ph]} \BibitemShut
  {NoStop}%
\bibitem [{Note1()}]{Note1}%
  \BibitemOpen
  \bibinfo {note} {See Appendix\protect \,\ref {app: formalism} for more
  rigorous treatment of the normalization.}\BibitemShut {Stop}%
\bibitem [{\citenamefont {{Gottesman}}(1997)}]{Gottesman1997S}%
  \BibitemOpen
  \bibfield  {author} {\bibinfo {author} {\bibfnamefont {D.}~\bibnamefont
  {{Gottesman}}},\ }\emph {\bibinfo {title} {{Stabilizer codes and quantum
  error correction}}},\ \href@noop {} {Ph.D. thesis},\ \bibinfo  {school}
  {California Institute of Technology} (\bibinfo {year} {1997})\BibitemShut
  {NoStop}%
\bibitem [{\citenamefont {{Gottesman}}(1998)}]{Gottesman1998Tquant-ph/9807006}%
  \BibitemOpen
  \bibfield  {author} {\bibinfo {author} {\bibfnamefont {D.}~\bibnamefont
  {{Gottesman}}},\ }\href {\doibase 10.48550/arXiv.quant-ph/9807006} {\bibfield
   {journal} {\bibinfo  {journal} {arXiv e-prints}\ ,\ \bibinfo {eid}
  {quant-ph/9807006}} (\bibinfo {year} {1998})},\ \Eprint
  {http://arxiv.org/abs/quant-ph/9807006} {arXiv:quant-ph/9807006 [quant-ph]}
  \BibitemShut {NoStop}%
\bibitem [{\citenamefont {{Ho}}\ and\ \citenamefont
  {{Abanin}}(2017)}]{Ho2017E1508.03784}%
  \BibitemOpen
  \bibfield  {author} {\bibinfo {author} {\bibfnamefont {W.~W.}\ \bibnamefont
  {{Ho}}}\ and\ \bibinfo {author} {\bibfnamefont {D.~A.}\ \bibnamefont
  {{Abanin}}},\ }\href {\doibase 10.1103/PhysRevB.95.094302} {\bibfield
  {journal} {\bibinfo  {journal} {\prb}\ }\textbf {\bibinfo {volume} {95}},\
  \bibinfo {eid} {094302} (\bibinfo {year} {2017})},\ \Eprint
  {http://arxiv.org/abs/1508.03784} {arXiv:1508.03784 [cond-mat.stat-mech]}
  \BibitemShut {NoStop}%
\bibitem [{\citenamefont {{Bohrdt}}\ \emph {et~al.}(2017)\citenamefont
  {{Bohrdt}}, \citenamefont {{Mendl}}, \citenamefont {{Endres}},\ and\
  \citenamefont {{Knap}}}]{Bohrdt2017S1612.02434}%
  \BibitemOpen
  \bibfield  {author} {\bibinfo {author} {\bibfnamefont {A.}~\bibnamefont
  {{Bohrdt}}}, \bibinfo {author} {\bibfnamefont {C.~B.}\ \bibnamefont
  {{Mendl}}}, \bibinfo {author} {\bibfnamefont {M.}~\bibnamefont {{Endres}}}, \
  and\ \bibinfo {author} {\bibfnamefont {M.}~\bibnamefont {{Knap}}},\ }\href
  {\doibase 10.1088/1367-2630/aa719b} {\bibfield  {journal} {\bibinfo
  {journal} {New Journal of Physics}\ }\textbf {\bibinfo {volume} {19}},\
  \bibinfo {eid} {063001} (\bibinfo {year} {2017})},\ \Eprint
  {http://arxiv.org/abs/1612.02434} {arXiv:1612.02434 [cond-mat.quant-gas]}
  \BibitemShut {NoStop}%
\bibitem [{\citenamefont {{Nahum}}\ \emph {et~al.}(2017)\citenamefont
  {{Nahum}}, \citenamefont {{Ruhman}}, \citenamefont {{Vijay}},\ and\
  \citenamefont {{Haah}}}]{Nahum2017Q1608.06950}%
  \BibitemOpen
  \bibfield  {author} {\bibinfo {author} {\bibfnamefont {A.}~\bibnamefont
  {{Nahum}}}, \bibinfo {author} {\bibfnamefont {J.}~\bibnamefont {{Ruhman}}},
  \bibinfo {author} {\bibfnamefont {S.}~\bibnamefont {{Vijay}}}, \ and\
  \bibinfo {author} {\bibfnamefont {J.}~\bibnamefont {{Haah}}},\ }\href
  {\doibase 10.1103/PhysRevX.7.031016} {\bibfield  {journal} {\bibinfo
  {journal} {Physical Review X}\ }\textbf {\bibinfo {volume} {7}},\ \bibinfo
  {eid} {031016} (\bibinfo {year} {2017})},\ \Eprint
  {http://arxiv.org/abs/1608.06950} {arXiv:1608.06950 [cond-mat.stat-mech]}
  \BibitemShut {NoStop}%
\bibitem [{\citenamefont {{Kukuljan}}\ \emph {et~al.}(2017)\citenamefont
  {{Kukuljan}}, \citenamefont {{Grozdanov}},\ and\ \citenamefont
  {{Prosen}}}]{Kukuljan2017W1701.09147}%
  \BibitemOpen
  \bibfield  {author} {\bibinfo {author} {\bibfnamefont {I.}~\bibnamefont
  {{Kukuljan}}}, \bibinfo {author} {\bibfnamefont {S.}~\bibnamefont
  {{Grozdanov}}}, \ and\ \bibinfo {author} {\bibfnamefont {T.}~\bibnamefont
  {{Prosen}}},\ }\href {\doibase 10.1103/PhysRevB.96.060301} {\bibfield
  {journal} {\bibinfo  {journal} {\prb}\ }\textbf {\bibinfo {volume} {96}},\
  \bibinfo {eid} {060301} (\bibinfo {year} {2017})},\ \Eprint
  {http://arxiv.org/abs/1701.09147} {arXiv:1701.09147 [cond-mat.stat-mech]}
  \BibitemShut {NoStop}%
\bibitem [{\citenamefont {{von Keyserlingk}}\ \emph {et~al.}(2018)\citenamefont
  {{von Keyserlingk}}, \citenamefont {{Rakovszky}}, \citenamefont
  {{Pollmann}},\ and\ \citenamefont
  {{Sondhi}}}]{von-Keyserlingk2018O1705.08910}%
  \BibitemOpen
  \bibfield  {author} {\bibinfo {author} {\bibfnamefont {C.~W.}\ \bibnamefont
  {{von Keyserlingk}}}, \bibinfo {author} {\bibfnamefont {T.}~\bibnamefont
  {{Rakovszky}}}, \bibinfo {author} {\bibfnamefont {F.}~\bibnamefont
  {{Pollmann}}}, \ and\ \bibinfo {author} {\bibfnamefont {S.~L.}\ \bibnamefont
  {{Sondhi}}},\ }\href {\doibase 10.1103/PhysRevX.8.021013} {\bibfield
  {journal} {\bibinfo  {journal} {Physical Review X}\ }\textbf {\bibinfo
  {volume} {8}},\ \bibinfo {eid} {021013} (\bibinfo {year} {2018})},\ \Eprint
  {http://arxiv.org/abs/1705.08910} {arXiv:1705.08910 [cond-mat.str-el]}
  \BibitemShut {NoStop}%
\bibitem [{\citenamefont {{Nahum}}\ \emph {et~al.}(2018)\citenamefont
  {{Nahum}}, \citenamefont {{Vijay}},\ and\ \citenamefont
  {{Haah}}}]{Nahum2018O1705.08975}%
  \BibitemOpen
  \bibfield  {author} {\bibinfo {author} {\bibfnamefont {A.}~\bibnamefont
  {{Nahum}}}, \bibinfo {author} {\bibfnamefont {S.}~\bibnamefont {{Vijay}}}, \
  and\ \bibinfo {author} {\bibfnamefont {J.}~\bibnamefont {{Haah}}},\ }\href
  {\doibase 10.1103/PhysRevX.8.021014} {\bibfield  {journal} {\bibinfo
  {journal} {Physical Review X}\ }\textbf {\bibinfo {volume} {8}},\ \bibinfo
  {eid} {021014} (\bibinfo {year} {2018})},\ \Eprint
  {http://arxiv.org/abs/1705.08975} {arXiv:1705.08975 [cond-mat.str-el]}
  \BibitemShut {NoStop}%
\bibitem [{\citenamefont {{Rakovszky}}\ \emph {et~al.}(2018)\citenamefont
  {{Rakovszky}}, \citenamefont {{Pollmann}},\ and\ \citenamefont {{von
  Keyserlingk}}}]{Rakovszky2018D1710.09827}%
  \BibitemOpen
  \bibfield  {author} {\bibinfo {author} {\bibfnamefont {T.}~\bibnamefont
  {{Rakovszky}}}, \bibinfo {author} {\bibfnamefont {F.}~\bibnamefont
  {{Pollmann}}}, \ and\ \bibinfo {author} {\bibfnamefont {C.~W.}\ \bibnamefont
  {{von Keyserlingk}}},\ }\href {\doibase 10.1103/PhysRevX.8.031058} {\bibfield
   {journal} {\bibinfo  {journal} {Physical Review X}\ }\textbf {\bibinfo
  {volume} {8}},\ \bibinfo {eid} {031058} (\bibinfo {year} {2018})},\ \Eprint
  {http://arxiv.org/abs/1710.09827} {arXiv:1710.09827 [cond-mat.stat-mech]}
  \BibitemShut {NoStop}%
\bibitem [{\citenamefont {{Khemani}}\ \emph {et~al.}(2018)\citenamefont
  {{Khemani}}, \citenamefont {{Vishwanath}},\ and\ \citenamefont
  {{Huse}}}]{Khemani2018O1710.09835}%
  \BibitemOpen
  \bibfield  {author} {\bibinfo {author} {\bibfnamefont {V.}~\bibnamefont
  {{Khemani}}}, \bibinfo {author} {\bibfnamefont {A.}~\bibnamefont
  {{Vishwanath}}}, \ and\ \bibinfo {author} {\bibfnamefont {D.~A.}\
  \bibnamefont {{Huse}}},\ }\href {\doibase 10.1103/PhysRevX.8.031057}
  {\bibfield  {journal} {\bibinfo  {journal} {Physical Review X}\ }\textbf
  {\bibinfo {volume} {8}},\ \bibinfo {eid} {031057} (\bibinfo {year} {2018})},\
  \Eprint {http://arxiv.org/abs/1710.09835} {arXiv:1710.09835
  [cond-mat.stat-mech]} \BibitemShut {NoStop}%
\bibitem [{\citenamefont {{Chan}}\ \emph {et~al.}(2018)\citenamefont {{Chan}},
  \citenamefont {{De Luca}},\ and\ \citenamefont
  {{Chalker}}}]{Chan2018S1712.06836}%
  \BibitemOpen
  \bibfield  {author} {\bibinfo {author} {\bibfnamefont {A.}~\bibnamefont
  {{Chan}}}, \bibinfo {author} {\bibfnamefont {A.}~\bibnamefont {{De Luca}}}, \
  and\ \bibinfo {author} {\bibfnamefont {J.~T.}\ \bibnamefont {{Chalker}}},\
  }\href {\doibase 10.1103/PhysRevX.8.041019} {\bibfield  {journal} {\bibinfo
  {journal} {Physical Review X}\ }\textbf {\bibinfo {volume} {8}},\ \bibinfo
  {eid} {041019} (\bibinfo {year} {2018})},\ \Eprint
  {http://arxiv.org/abs/1712.06836} {arXiv:1712.06836 [cond-mat.stat-mech]}
  \BibitemShut {NoStop}%
\bibitem [{\citenamefont {{Zhou}}\ and\ \citenamefont
  {{Chen}}(2019)}]{Zhou2019O1805.09307}%
  \BibitemOpen
  \bibfield  {author} {\bibinfo {author} {\bibfnamefont {T.}~\bibnamefont
  {{Zhou}}}\ and\ \bibinfo {author} {\bibfnamefont {X.}~\bibnamefont
  {{Chen}}},\ }\href {\doibase 10.1103/PhysRevE.99.052212} {\bibfield
  {journal} {\bibinfo  {journal} {\pre}\ }\textbf {\bibinfo {volume} {99}},\
  \bibinfo {eid} {052212} (\bibinfo {year} {2019})},\ \Eprint
  {http://arxiv.org/abs/1805.09307} {arXiv:1805.09307 [cond-mat.str-el]}
  \BibitemShut {NoStop}%
\bibitem [{\citenamefont {{Zhou}}\ and\ \citenamefont
  {{Nahum}}(2019)}]{Zhou2019E1804.09737}%
  \BibitemOpen
  \bibfield  {author} {\bibinfo {author} {\bibfnamefont {T.}~\bibnamefont
  {{Zhou}}}\ and\ \bibinfo {author} {\bibfnamefont {A.}~\bibnamefont
  {{Nahum}}},\ }\href {\doibase 10.1103/PhysRevB.99.174205} {\bibfield
  {journal} {\bibinfo  {journal} {\prb}\ }\textbf {\bibinfo {volume} {99}},\
  \bibinfo {eid} {174205} (\bibinfo {year} {2019})},\ \Eprint
  {http://arxiv.org/abs/1804.09737} {arXiv:1804.09737 [cond-mat.stat-mech]}
  \BibitemShut {NoStop}%
\bibitem [{\citenamefont {{Xu}}\ and\ \citenamefont
  {{Swingle}}(2019)}]{Xu2019L1805.05376}%
  \BibitemOpen
  \bibfield  {author} {\bibinfo {author} {\bibfnamefont {S.}~\bibnamefont
  {{Xu}}}\ and\ \bibinfo {author} {\bibfnamefont {B.}~\bibnamefont
  {{Swingle}}},\ }\href {\doibase 10.1103/PhysRevX.9.031048} {\bibfield
  {journal} {\bibinfo  {journal} {Physical Review X}\ }\textbf {\bibinfo
  {volume} {9}},\ \bibinfo {eid} {031048} (\bibinfo {year} {2019})},\ \Eprint
  {http://arxiv.org/abs/1805.05376} {arXiv:1805.05376 [cond-mat.str-el]}
  \BibitemShut {NoStop}%
\bibitem [{\citenamefont {{Chen}}\ and\ \citenamefont
  {{Zhou}}(2019)}]{Chen2019Q1808.09812}%
  \BibitemOpen
  \bibfield  {author} {\bibinfo {author} {\bibfnamefont {X.}~\bibnamefont
  {{Chen}}}\ and\ \bibinfo {author} {\bibfnamefont {T.}~\bibnamefont
  {{Zhou}}},\ }\href {\doibase 10.1103/PhysRevB.100.064305} {\bibfield
  {journal} {\bibinfo  {journal} {\prb}\ }\textbf {\bibinfo {volume} {100}},\
  \bibinfo {eid} {064305} (\bibinfo {year} {2019})},\ \Eprint
  {http://arxiv.org/abs/1808.09812} {arXiv:1808.09812 [cond-mat.stat-mech]}
  \BibitemShut {NoStop}%
\bibitem [{\citenamefont {{Parker}}\ \emph {et~al.}(2019)\citenamefont
  {{Parker}}, \citenamefont {{Cao}}, \citenamefont {{Avdoshkin}}, \citenamefont
  {{Scaffidi}},\ and\ \citenamefont {{Altman}}}]{Parker2019A1812.08657}%
  \BibitemOpen
  \bibfield  {author} {\bibinfo {author} {\bibfnamefont {D.~E.}\ \bibnamefont
  {{Parker}}}, \bibinfo {author} {\bibfnamefont {X.}~\bibnamefont {{Cao}}},
  \bibinfo {author} {\bibfnamefont {A.}~\bibnamefont {{Avdoshkin}}}, \bibinfo
  {author} {\bibfnamefont {T.}~\bibnamefont {{Scaffidi}}}, \ and\ \bibinfo
  {author} {\bibfnamefont {E.}~\bibnamefont {{Altman}}},\ }\href {\doibase
  10.1103/PhysRevX.9.041017} {\bibfield  {journal} {\bibinfo  {journal}
  {Physical Review X}\ }\textbf {\bibinfo {volume} {9}},\ \bibinfo {eid}
  {041017} (\bibinfo {year} {2019})},\ \Eprint
  {http://arxiv.org/abs/1812.08657} {arXiv:1812.08657 [cond-mat.stat-mech]}
  \BibitemShut {NoStop}%
\bibitem [{\citenamefont {{Qi}}\ \emph {et~al.}(2019)\citenamefont {{Qi}},
  \citenamefont {{Davis}}, \citenamefont {{Periwal}},\ and\ \citenamefont
  {{Schleier-Smith}}}]{Qi2019M1906.00524}%
  \BibitemOpen
  \bibfield  {author} {\bibinfo {author} {\bibfnamefont {X.-L.}\ \bibnamefont
  {{Qi}}}, \bibinfo {author} {\bibfnamefont {E.~J.}\ \bibnamefont {{Davis}}},
  \bibinfo {author} {\bibfnamefont {A.}~\bibnamefont {{Periwal}}}, \ and\
  \bibinfo {author} {\bibfnamefont {M.}~\bibnamefont {{Schleier-Smith}}},\
  }\href {\doibase 10.48550/arXiv.1906.00524} {\bibfield  {journal} {\bibinfo
  {journal} {arXiv e-prints}\ ,\ \bibinfo {eid} {arXiv:1906.00524}} (\bibinfo
  {year} {2019})},\ \Eprint {http://arxiv.org/abs/1906.00524} {arXiv:1906.00524
  [quant-ph]} \BibitemShut {NoStop}%
\bibitem [{\citenamefont {{Kuo}}\ \emph {et~al.}(2020)\citenamefont {{Kuo}},
  \citenamefont {{Akhtar}}, \citenamefont {{Arovas}},\ and\ \citenamefont
  {{You}}}]{Kuo2020M1910.11351}%
  \BibitemOpen
  \bibfield  {author} {\bibinfo {author} {\bibfnamefont {W.-T.}\ \bibnamefont
  {{Kuo}}}, \bibinfo {author} {\bibfnamefont {A.~A.}\ \bibnamefont {{Akhtar}}},
  \bibinfo {author} {\bibfnamefont {D.~P.}\ \bibnamefont {{Arovas}}}, \ and\
  \bibinfo {author} {\bibfnamefont {Y.-Z.}\ \bibnamefont {{You}}},\ }\href
  {\doibase 10.1103/PhysRevB.101.224202} {\bibfield  {journal} {\bibinfo
  {journal} {\prb}\ }\textbf {\bibinfo {volume} {101}},\ \bibinfo {eid}
  {224202} (\bibinfo {year} {2020})},\ \Eprint
  {http://arxiv.org/abs/1910.11351} {arXiv:1910.11351 [cond-mat.dis-nn]}
  \BibitemShut {NoStop}%
\bibitem [{\citenamefont {{Akhtar}}\ and\ \citenamefont
  {{You}}(2020)}]{Akhtar2020M2006.08797}%
  \BibitemOpen
  \bibfield  {author} {\bibinfo {author} {\bibfnamefont {A.~A.}\ \bibnamefont
  {{Akhtar}}}\ and\ \bibinfo {author} {\bibfnamefont {Y.-Z.}\ \bibnamefont
  {{You}}},\ }\href {\doibase 10.1103/PhysRevB.102.134203} {\bibfield
  {journal} {\bibinfo  {journal} {\prb}\ }\textbf {\bibinfo {volume} {102}},\
  \bibinfo {eid} {134203} (\bibinfo {year} {2020})},\ \Eprint
  {http://arxiv.org/abs/2006.08797} {arXiv:2006.08797 [cond-mat.dis-nn]}
  \BibitemShut {NoStop}%
\bibitem [{Note2()}]{Note2}%
  \BibitemOpen
  \bibinfo {note} {See Appendix\protect \,\ref {app: formalism} for a brief
  review of the Markov evolution of Pauli weights.}\BibitemShut {Stop}%
\bibitem [{Note3()}]{Note3}%
  \BibitemOpen
  \bibinfo {note} {Strictly speaking, the classical snapshot Pauli weight
  dynamics is not Markovian due to the normalization factor on the denominator
  of Eq.\protect \,\protect \textup {\hbox {\mathsurround \z@ \protect
  \normalfont (\ignorespaces \ref {eq: def sigma}\unskip \@@italiccorr )}}. We
  are taking a Markovian approximation, that enables us to make
  thermodynamic-limit estimation of the shadow norm. The approximation is
  expected to work away from the measurement-induced critical point when the
  critical fluctuation is suppressed.}\BibitemShut {Stop}%
\bibitem [{\citenamefont {{Greenberger}}\ \emph {et~al.}(2007)\citenamefont
  {{Greenberger}}, \citenamefont {{Horne}},\ and\ \citenamefont
  {{Zeilinger}}}]{Greenberger2007G0712.0921}%
  \BibitemOpen
  \bibfield  {author} {\bibinfo {author} {\bibfnamefont {D.~M.}\ \bibnamefont
  {{Greenberger}}}, \bibinfo {author} {\bibfnamefont {M.~A.}\ \bibnamefont
  {{Horne}}}, \ and\ \bibinfo {author} {\bibfnamefont {A.}~\bibnamefont
  {{Zeilinger}}},\ }\href {\doibase 10.48550/arXiv.0712.0921} {\bibfield
  {journal} {\bibinfo  {journal} {arXiv e-prints}\ ,\ \bibinfo {eid}
  {arXiv:0712.0921}} (\bibinfo {year} {2007})},\ \Eprint
  {http://arxiv.org/abs/0712.0921} {arXiv:0712.0921 [quant-ph]} \BibitemShut
  {NoStop}%
\bibitem [{Note4()}]{Note4}%
  \BibitemOpen
  \bibinfo {note} {See Appendix\protect \,\ref {app: statmech} for a
  statistical mechanical model interpretation for the scaling
  behavior.}\BibitemShut {Stop}%
\bibitem [{Note5()}]{Note5}%
  \BibitemOpen
  \bibinfo {note} {See Appendix\protect \,\ref {app: toy} for a more detailed
  quantitative analysis using toy models.}\BibitemShut {Stop}%
\bibitem [{\citenamefont {{Li}}\ \emph {et~al.}(2023)\citenamefont {{Li}},
  \citenamefont {{Zou}}, \citenamefont {{Glorioso}}, \citenamefont {{Altman}},\
  and\ \citenamefont {{Fisher}}}]{Li2023C2209.00609}%
  \BibitemOpen
  \bibfield  {author} {\bibinfo {author} {\bibfnamefont {Y.}~\bibnamefont
  {{Li}}}, \bibinfo {author} {\bibfnamefont {Y.}~\bibnamefont {{Zou}}},
  \bibinfo {author} {\bibfnamefont {P.}~\bibnamefont {{Glorioso}}}, \bibinfo
  {author} {\bibfnamefont {E.}~\bibnamefont {{Altman}}}, \ and\ \bibinfo
  {author} {\bibfnamefont {M.~P.~A.}\ \bibnamefont {{Fisher}}},\ }\href
  {\doibase 10.1103/PhysRevLett.130.220404} {\bibfield  {journal} {\bibinfo
  {journal} {\prl}\ }\textbf {\bibinfo {volume} {130}},\ \bibinfo {eid}
  {220404} (\bibinfo {year} {2023})},\ \Eprint
  {http://arxiv.org/abs/2209.00609} {arXiv:2209.00609 [quant-ph]} \BibitemShut
  {NoStop}%
\bibitem [{\citenamefont {{Ippoliti}}\ and\ \citenamefont
  {{Khemani}}(2023)}]{Ippoliti2023L2307.15011}%
  \BibitemOpen
  \bibfield  {author} {\bibinfo {author} {\bibfnamefont {M.}~\bibnamefont
  {{Ippoliti}}}\ and\ \bibinfo {author} {\bibfnamefont {V.}~\bibnamefont
  {{Khemani}}},\ }\href {\doibase 10.48550/arXiv.2307.15011} {\bibfield
  {journal} {\bibinfo  {journal} {arXiv e-prints}\ ,\ \bibinfo {eid}
  {arXiv:2307.15011}} (\bibinfo {year} {2023})},\ \Eprint
  {http://arxiv.org/abs/2307.15011} {arXiv:2307.15011 [quant-ph]} \BibitemShut
  {NoStop}%
\end{thebibliography}%

\onecolumngrid
\newpage
\appendix
\section{General Formalism}\label{app: formalism}

\subsection{Prior and Posterior Ensembles of Classical Snapshots}

Assume the Krause operator (projection operator) $K_{\vect{b}|\scC}$ describes a projective measurement implemented by a quantum circuit $\scC$ (with specific gate and observable choices) and resulted in the measurement outcomes $\vect{b}$, such that the probability to observe $\vect{b}$ on a state $\rho$ is given by
\eq{\label{eq: posterior conditional}
p(\vect{b}|\rho,\scC)=\Tr(K_{\vect{b}|\scC}\rho K_{\vect{b}|\scC}^\dagger).}
If we have no knowledge about the state $\rho$, we should assume $\rho=\id/(\Tr \id)$ to be maximally mixed, where $\id$ stands for the identity operator in the Hilbert space and $\Tr\id$ is effectively the Hilbert space dimension. In this 
limit, \eqnref{eq: posterior conditional} reduces to
\eq{\label{eq: prior conditional}
p(\vect{b}|\scC):=p(\vect{b}|\rho=\tfrac{\id}{\Tr\id},\scC)=\frac{1}{\Tr\id}\Tr(K_{\vect{b}|\scC} K_{\vect{b}|\scC}^\dagger).}
The operator $K_{\vect{b}|\scC}^\dagger K_{\vect{b}|\scC}$ is a Hermitian and positive semi-definite operator, which motivates us to further normalize it to make it a state (a density matrix)
\eq{\label{eq: app def sigma}
\sigma_{\vect{b}|C}=\frac{K_{\vect{b}|\scC}^\dagger K_{\vect{b}|\scC}}{\Tr(K_{\vect{b}|\scC}^\dagger K_{\vect{b}|\scC})}.}
We will call $\sigma_{\vect{b}|C}$ a \emph{classical snapshot} state. It has an important property that
\eq{\label{eq: Bayes ratio}
\Tr(\sigma_{\vect{b}|C}\rho)=\frac{\Tr(K_{\vect{b}|\scC}^\dagger K_{\vect{b}|\scC}\rho)}{\Tr(K_{\vect{b}|\scC}^\dagger K_{\vect{b}|\scC})}=\frac{1}{\Tr\id}\frac{p(\vect{b}|\rho,\scC)}{p(\vect{b}|\scC)}.}

Assume that the measurement circuit $\scC$ is drawn from some random ensemble with probability $p(\scC)$, we can define two random state ensembles for the classical snapshots:
\begin{itemize}
\item the \emph{prior} snapshot ensemble (with no knowledge about $\rho$)
\eq{
\scE_\sigma=\{\sigma_{\vect{b}|\scC}|\;(\vect{b},\scC)\sim p(\vect{b},\scC):=p(\vect{b}|\scC)p(\scC)\},
}
\item the \emph{posterior} snapshot ensemble (given the knowledge about $\rho$)
\eq{
\scE_{\sigma|\rho}=\{\sigma_{\vect{b}|\scC}|\;(\vect{b},\scC)\sim p(\vect{b},\scC|\rho):=p(\vect{b}|\rho,\scC)p(\scC)\}.
}
\end{itemize}
This means the ensemble averages are defined as
\eqs{\E_{\sigma\in\scE_\sigma}f(\sigma)&:=\sum_{\vect{b},\scC}f(\sigma_{\vect{b}|\scC})p(\vect{b}|\scC)p(\scC),\\
\E_{\sigma\in\scE_{\sigma|\rho}}f(\sigma)&:=\sum_{\vect{b},\scC}f(\sigma_{\vect{b}|\scC})p(\vect{b}|\rho,\scC)p(\scC),}
where $f(\sigma)$ stands for any function of $\sigma$, $p(\vect{b}|\scC)$ and $p(\vect{b}|\rho,\scC)$ are given by \eqnref{eq: prior conditional} and \eqnref{eq: posterior conditional} respectively. Then \eqnref{eq: Bayes ratio} implies that the posterior ensemble average can be expressed as a prior ensemble average as
\eq{\label{eq: posterior-prior relation}\E_{\sigma\in\scE_{\sigma|\rho}}f(\sigma)=(\Tr\id)\E_{\sigma\in\scE_{\sigma}}f(\sigma)\Tr(\sigma\rho).}

\subsection{Review of Classical Shadow Tomography}

Classical shadow tomography is an efficient approach to extracting information about an unknown quantum state $\rho$ by repeated randomized measurements. The idea of randomized measurement is to sample the measurement circuit $\scC$ (both its structure and its gate choices) from a probability distribution $p(\scC)$, perform the measurement on the state $\rho$ and collect the measurement outcomes $\vect{b}$. The randomized measurement effectively converts the quantum state $\rho$ into a collection of classical snapshots in the posterior ensemble $\scE_{\sigma|\rho}$.

The data acquisition procedure can be formulated as a quantum channel $\scM$, called the \emph{measurement channel}, which maps the initial quantum state $\rho$ to the expectation of classical snapshot states over the posterior snapshot ensemble
\eq{\label{eq: def M}
\scM(\rho):=\E_{\sigma\in\scE_{\sigma|\rho}}\sigma=(\Tr\id)\E_{\sigma\in\scE_{\sigma}}\sigma\Tr(\sigma\rho).
}
Here we have used \eqnref{eq: posterior-prior relation}. Suppose the randomized measurement scheme is topographically complete, the measurement channel $\scM$ will be invertible. Its inverse is called the \emph{reconstruction map}, denoted as $\scM^{-1}$ (albeit $\scM^{-1}$ might not be a physical channel), such that the quantum state $\rho$ can be reconstructed from the classical snapshots by $\rho=\E_{\sigma\in\scE_{\sigma|\rho}}\scM^{-1}(\sigma)$. This also provides the means to predict the expectation value of any observable $O$ on the state $\rho$ as
\eq{\label{eq: def <O>}
\langle O\rangle:=\Tr (O\rho)=\E_{\sigma\in\scE_{\sigma|\rho}}\Tr(\scM^{-1}(O)\sigma).
}
Note that $\Tr(O\scM^{-1}(\sigma))=\Tr(\scM^{-1}(O)\sigma)$ due to the self-adjoint property of the measurement channel $\scM$ as well as the reconstruction map $\scM^{-1}$.

In practice, the expectation $\E_{\sigma\in\scE_{\sigma|\rho}}$ is often estimated by the median-of-means over a finite number of classical snapshots collected from experiments. Due to the statistical fluctuation of finite samples, the estimation value of an observable $O$ will fluctuate around its true expectation value $\langle O\rangle$ with a typical variance that scales with the sample number $M$ as $\var O\sim \Vert O\Vert_{\scE_\sigma}^2/M$ following the law of large numbers. The coefficient $\Vert O\Vert_{\scE_\sigma}^2$ is the \emph{locally-scrambled shadow norm}, defined as
\eq{\label{eq: def shadow norm}
\Vert O\Vert_{\scE_\sigma}^2:=\E_{\sigma\in\scE_{\sigma}}(\Tr \scM^{-1}(O)\sigma)^2=\frac{\Tr(O\scM^{-1}(O))}{\Tr\id}.
}
By definition, the identity operator $O=\id$ always have unit shadow norm, i.e., $\Vert \id\Vert_{\scE_\sigma}^2\equiv 1$. For traceless observable $O$, the locally-scrambled shadow norm $\Vert O\Vert_{\scE_\sigma}^2$ quantifies the number $M$ of samples needed to control the estimation variances $\var O\lesssim\delta^2$ below a desired level set by a small $\delta$, as $M\sim \Vert O\Vert_{\scE_\sigma}^2/\delta^2$. Therefore, the locally-scrambled shadow norm measures the \emph{sample complexity} for classical shadow tomography to predict the observable $O$ based on the randomized measurement scheme characterized by $\scE_\sigma$. 

\subsection{Locally Scrambled Ensembles and Pauli Basis Approach}

From the aforementioned general formulation, it is evident that the measurement channel $\scM$ and its inverse $\scM^{-1}$ hold a central role in classical shadow tomography. Computing these for generic randomized measurement schemes is a complex task, and no polynomially scalable algorithm currently exists. Nevertheless, progress has been made in the context of locally scrambled (or Pauli-twirled) measurements, which are randomized measurements insensitive to local basis choices.

More specifically, let $V\in\U(2)^N$ be a product of single-qubit unitary operators, decomposed as $V=\bigotimes_i V_i$, with  $V_i\in\U(2)$ on every qubit $i$. The product operator $V$ represents independent local basis transformations. Let $\scU_V(\rho):=V\rho V^\dagger$ be the unitary channel that implements the unitary transformation $V$. A random state ensemble $\scE_\rho=\{\rho|\rho\sim p(\rho)\}$ is said to be locally scrambled, if and only if $p(\rho)=p(\scU_V(\rho))$ for any $V\in\U(2)^N$. A random channel ensemble $\scE_\scK=\{\scK|\scK\sim p(\scK)\}$ is considered locally scrambled, if and only if $p(\scK)=p(\scU_V\circ\scK\circ\scU_{V'})$ for any $V, V'\in\U(2)^N$. A randomized measurement scheme is locally scrambled if its corresponding prior snapshot ensemble $\scE_\sigma$ (as a random state ensemble) is locally scrambled. For locally scrambled randomized measurements, the associated $\scM$ and $\scM^{-1}$ exhibit simple forms in the Pauli basis and can be computed efficiently. This conclusion remains valid when the scrambling condition is relaxed to $V\in C_1^N$, where $C_1$ denotes the single-qubit Clifford group, which is applicable to the cases of random Clifford gates with random Pauli measurements.

To elucidate this approach, it is more convenient to employ the Choi representation, wherein each quantum operator $O$ is viewed as a super-state $\Ket{O}$:
\eq{
\Ket{O}:=\frac{1}{\sqrt{\Tr\id}}\sum_{a}\ket{a}\otimes O\ket{a},
}
where $a$ labels a complete set of orthonormal basis in the Hilbert space $\scH$ (of ket states). All Hermitian operators in $\scH$ span an operator space, denoted $\scL(\scH)$, in which inner product between two operators $A,B\in\scL(\scH)$ is defined as
\eq{
\Braket{A}{B}:=\frac{\Tr A^\dagger B}{\Tr \id}.
}
Let $P=\prod_iP_i$ represent a (multi-qubit) Pauli operator, constructed as a product of single-qubit Pauli operators, with $P_i\in{I,X,Y,Z}$ acting on the $i$th qubit. All Pauli operators form a complete set of orthonormal basis for the Hermitian operator space $\scL(\scH)$, as $\Braket{P}{P'}=\delta_{PP'}$. Correspondingly, for any quantum channel $\scK$ that maps a state $\rho$ to another state $\rho'=\scK(\rho)$, there exists a corresponding super-operator $\hat{\scK}\in\text{Hom}(\scL(\scH), \scL(\scH))$ that maps their Choi representations accordingly, as $\Ket{\rho'}=\hat{\scK}\Ket{\rho}$. 

Utilizing the Choi representation of quantum channel, any locally scrambled measurement channel $\scM$ defined in \eqnref{eq: def M} can be written as a super-operator $\hat{\scM}$ that acts on any state $\rho$ as
\eq{\hat{\scM}\Ket{\rho}=(\Tr\id)^2\E_{\sigma\in\scE_\sigma}\Ket{\sigma}\Braket{\sigma}{\rho},}
which indicates that $\hat{\scM}=(\Tr\id)^2\E_{\sigma\in\scE_\sigma}\Ket{\sigma}\Bra{\sigma}$. Expand in Pauli basis, we generally have 
\eq{\label{eq: M general Pauli}
\hat{\scM}=(\Tr\id)^2\sum_{P,P'}\Ket{P}\Big(\E_{\sigma\in\scE_\sigma}\Braket{P}{\sigma}\Braket{\sigma}{P'}\Big)\Bra{P'}.}
However, if we assume $\scE_\sigma$ to be a locally scrambled state ensemble, the local scrambling property requires
\eq{\E_{\sigma\in\scE_\sigma}\Braket{P}{\sigma}\Braket{\sigma}{P'}=\delta_{PP'} \E_{\sigma\in\scE_\sigma}\Braket{P}{\sigma}\Braket{\sigma}{P}=\delta_{PP'} \E_{\sigma\in\scE_\sigma}\Big(\frac{\Tr P\sigma}{\Tr\id}\Big)^2.}
Therefore, the measurement channel \eqnref{eq: M general Pauli} becomes diagonal in the Pauli basis:
\eq{
\hat{\scM}=\sum_{P}\Ket{P}w_{\scE_\sigma}(P)\Bra{P},
}
where $w_{\scE_\sigma}(P)$ denotes the average \emph{Pauli weight} of classical snapshots $\sigma$ in prior snapshot ensemble $\scE_\sigma$, defined as
\eq{
w_{\scE_\sigma}(P):=\E_{\sigma\in\scE_\sigma}(\Tr P\sigma)^2.
}
This definition captures an essential statistical feature of the randomized measurement scheme: $w_{\scE_\sigma}(P)$ is the probability for the Pauli operator $P$ to appear diagonal in the measurement basis and be directly observed in a single random realization of the measurement protocol.

Since $\hat{\scM}$ is diagonal in the Pauli basis, its inverse is simply given by
\eq{
\hat{\scM}^{-1}=\sum_{P}\Ket{P}\frac{1}{w_{\scE_\sigma}(P)}\Bra{P},
}
or, more explicitly, in terms of the reconstruction map on any operator $O$ as
\eq{\label{eq: sol Minv}
\scM^{-1}(O)=\sum_{P} \frac{\Tr(O P)}{w_{\scE_\sigma}(P)\Tr \id} P.
}
Plugging \eqnref{eq: sol Minv} into \eqnref{eq: def <O>} and \eqnref{eq: def shadow norm} enables us to estimate the expectation value $\langle O\rangle$ and study the shadow norm $\Vert O\Vert_{\scE_\sigma}^2$ for any observable $O$. In particular, for Pauli observable $P$, we have
\eqs{
\langle P\rangle&=\frac{1}{w_{\scE_\sigma}(P)}\E_{\sigma\in\scE_{\sigma|\rho}}\Tr(P\sigma), \\
\Vert P\Vert_{\scE_\sigma}^2&=\frac{1}{w_{\scE_\sigma}(P)}.
}
This essentially allows us to decode the quantum information in the initial state $\rho$ from the measurement outcomes gathered from a hybrid quantum circuit and to investigate the sample complexity for decoding various observables at different measurement rates. 

\subsection{Evolution of Pauli Weights through Locally Scrambled Channels}

The central challenge now is to compute the average Pauli weight $w_{\scE_\sigma}(P)$ of the prior snapshot ensemble $\scE_\sigma$. In general, this remains a difficult problem. However, suppose the state ensemble $\scE_\sigma$ is constructed by applying locally scrambled elementary quantum channels to locally scrambled simple initial states. In that case, its Pauli weight can be calculated using Markov dynamics, which is a more tractable approach.

To derive the dynamics of Pauli weights under locally scrambled quantum dynamics, we first introduce a set of \emph{region basis} states $\Ket{A}$ in the doubled Hermitian operator space $\scL(\scH)^{2}$, defined by
\eq{
\Ket{A}=\frac{1}{\sqrt{|\scP_A|}}\sum_{P\in\scP_A}\Ket{P}\otimes\Ket{P},
}
where $A$ denotes a subset of qubits and $\scP_A=\{P|\supp P = A\}$ denotes all Pauli operators supported exactly in region $A$. $|\scP_A|=3^{|A|}$ is the cardinality of $\scP_A$ and $|A|$ denotes the number of qubits in $A$. The significance of $\Ket{A}$ lies in its invariance under doubled local basis transformation, i.e., $\hat{\scU}_V\otimes \hat{\scU}_V \Ket{A}=\Ket{A}$ for any $V\in\U(2)^N$. In fact, all states $\Ket{A}$ form a complete set of orthonormal basis that spans the invariant subspace of $\scL(\scH)^{2}$ under the doubled local scrambling $\hat{\scU}_V\otimes \hat{\scU}_V$.

Given a locally scrambled state ensemble $\scE_\rho$ and a locally scrambled channel ensemble $\scE_\scK$, by definition, for any local basis transformation $V=\bigotimes_iV_i\in\U(2)^N$ (or its corresponding unitary channel $\scU_V(\rho)=V\rho V^\dagger$), we have the following symmetry requirements 
\eqs{
\hat{\scU}_V^{\otimes 2}\E_{\rho\in\scE_\rho}\Ket{\rho}^{\otimes 2}&=\E_{\rho\in\scE_\rho}\Ket{\rho}^{\otimes 2},\\
\hat{\scU}_V^{\otimes 2}\Big(\E_{\scK\in\scE_\scK}\hat{\scK}^{\otimes 2}\Big)\hat{\scU}_{V^\dagger}^{\otimes 2}&=\E_{\scK\in\scE_\scK}\hat{\scK}^{\otimes2}.
}
These manifest the locally scrambled properties on the 2nd-moment level. Given these symmetry requirements, the 2nd-moment locally scrambled random states (or channels) can and only need to be represented in the symmetric subspace spanned by the region basis as
\eqs{\label{eq: regional dec}
\E_{\rho\in\scE_\rho}\Ket{\rho}^{\otimes 2}&=\frac{1}{(\Tr\id)^2}\sum_{A} \bar{w}_{\scE_\rho}(A)\Ket{A},\\
\E_{\scK\in\scE_\scK}\hat{\scK}^{\otimes 2}&=\sum_{A,A'} \Ket{A}\bar{w}_{\scE_\scK}(A,A')\Bra{A'},\\
}
where the linear combination coefficients $\bar{w}_\rho(A)$ and $\bar{w}_\scK(A,A')$ are defined as
\eqs{
\bar{w}_{\scE_\rho}(A)&:=(\Tr\id)^2\E_{\rho\in\scE_\rho}\Bra{A}\Ket{\rho}^{\otimes 2},\\
\bar{w}_{\scE_\scK}(A,A')&:=\E_{\scK\in\scE_\scK}\Bra{A}\hat{\scK}^{\otimes 2}\Ket{A'},\\
}
which will be called the regional Pauli weights of the locally scrambled ensembles.

Locally scrambled quantum dynamics involve passing locally scrambled random states $\rho\in\scE_\rho$ through locally scrambled random channels $\scK\in\scE_\scK$ such that the resulting states $\scK(\rho)$ still form a locally scrambled random state ensemble $\scE_{\scK(\rho)}$ defined as:
\eq{
\scE_\rho\to\scE_{\scK(\rho)}:=\{\scK(\rho)|\rho\in\scE_\rho,\scK\in\scE_\scK\}.
}
Under such evolution, the 2nd-moment evolves as
\eq{
\E_{\scK(\rho)\in\scE_{\scK(\rho)}}\Ket{\scK(\rho)}^{\otimes 2}=\E_{\scK\in\scE_\scK}\hat{\scK}^{\otimes 2}\E_{\rho\in\scE_\rho}\Ket{\rho}^{\otimes 2}.
}
By decomposing these 2nd-moment state and channel expectations in the region basis according to \eqnref{eq: regional dec}, we obtain the following evolution equation for the regional Pauli weights:
\eq{\label{eq: wA evolve}
\bar{w}_{\scE_{\scK(\rho)}}(A)=\sum_{A'}\bar{w}_{\scE_\scK}(A,A')\bar{w}_{\scE_\rho}(A').
}
Suppose a locally scrambled state ensemble is obtained from random quantum dynamics comprised of locally scrambled simple channels, with computable regional Pauli weights. In that case, we can use \eqnref{eq: wA evolve} to trace the evolution of the regional Pauli weights of the state ensemble step-by-step, eventually inferring the regional Pauli weights of the final state ensemble. This approach enables us to systematically calculate the Pauli weights of the prior snapshot ensemble $\scE_\sigma$, which is the key for classical shadow tomography.

Finally, the regional Pauli weights and the ordinary Pauli weights are connected by the following relations, given $P$ and $P'$ are Pauli operators supported on regions $A$ and $A'$, respectively,
\eqs{\label{eq: def Pauli weight}
w_{\scE_\rho}(P)&=\frac{\bar{w}_{\scE_\rho}(A)}{\sqrt{3}^{|A|}}=\E_{\rho\in\scE_\rho}(\Tr P \rho)^2,\\
w_{\scE_\scK}(P,P')&=\frac{\bar{w}_{\scE_\scK}(A,A')}{\sqrt{3}^{|A|+|A'|}}=\E_{\scK\in\scE_\scK}\Big(\frac{\Tr(P\scK(P'))}{\Tr\id}\Big)^2.
}
Switching back to the Pauli basis, \eqnref{eq: wA evolve} reduces to a similar form
\eq{\label{eq: wP evolve}
w_{\scE_{\scK(\rho)}}(P)=\sum_{P'}w_{\scE_\scK}(P,P')w_{\scE_\rho}(P').
}
Based on these formulae, we can compute the Pauli weights for a few locally scrambled states and channels, which will be useful for analyzing the hybrid quantum circuit classical shadow tomography.
\begin{itemize}

\item Locally scrambled state ensembles $\scE_\rho$

\begin{itemize}

\item Pure random product states $\scE_\rho=\{\bigotimes_i\ket{\psi_i}\bra{\psi_i}\;|\;\ket{\psi_i}\in\scH_i\}$
\eq{w_{\scE_\rho}(P)=\frac{1}{3^{|\supp P|}}.}

\item Pure Page state over $N$ qubits $\scE_\rho=\{\ket{\psi}\bra{\psi}\;|\;\ket{\psi}\in\scH=\bigotimes_i\scH_i\}$
\eq{w_{\scE_\rho}(P)=\left\{
\begin{array}{ll}
1 & P=\id,\\
\frac{1}{2^N+1} & \text{otherwise}.
\end{array}\right.}

\item Maximally mixed state $\scE_\rho=\{\id/(\Tr\id)\}$
\eq{\label{eq: w maxmix}
w_{\scE_\rho}(P)=\delta_{P,\id}=\left\{
\begin{array}{ll}
1 & P=\id,\\
0 & \text{otherwise}.
\end{array}\right.}

\end{itemize}

\item Locally scrambled channel ensembles $\scE_{\scK}$

\begin{itemize}

\item Local scrambling channel $\scE_\scK=\{\scU_V|\scU_V(\rho)=V\rho V^\dagger, V\in\U(2)^N\}$
\eq{w_{\scE_\scK}(P,P')=\frac{\delta_{\supp P, \supp P'}}{3^{|\supp P|}}.
}

\item Global scrambling channel among $N$ qubits $\scE_\scK=\{\scU_U|\scU_U(\rho)=U\rho U^\dagger, U\in\U(2^N)\}$
\eq{w_{\scE_\scK}(P,P')=\delta_{P,\id}\delta_{P',\id}+\frac{(1-\delta_{P,\id})(1-\delta_{P',\id})}{4^N-1}.
}

\item Local random projective measurement $\scE_\scK=\{\scM|\scM(\rho_i)=\ket{\psi_i}\bra{\phi_i}\rho_i\ket{\phi_i}\bra{\psi_i}, \ket{\phi_i},\ket{\psi_i}\in\scH_i\}$
\eq{w_{\scE_\scK}(P,P')=\frac{1}{3^{|\supp P|+|\supp P'|}}.}
\end{itemize}
\end{itemize}

\subsection{Application to the Prior Snapshot Ensemble}

According to \eqnref{eq: app def sigma}, the classical snapshot state $\sigma$ takes the form of
\eq{\sigma=\frac{K^\dagger K}{\Tr(K^\dagger K)}.}
The state can be interpreted as mapping the maximal mixed state $\rho_\id:=\id/(\Tr\id)$ by a quantum channel $\scK$, followed by a normalization,
\eq{\sigma=\frac{\scK(\rho_\id)}{\Tr \scK(\rho_\id)},}
where the quantum channel $\scK$ is defined by the Krause operator as $\scK(\rho)=K^\dagger \rho K$. The random state ensemble of $\sigma$ can be considered as derived from the random channel ensemble of $\scK$,
\eq{\scE_\sigma=\Big\{\frac{\scK(\rho_\id)}{\Tr \scK(\rho_\id)}\Big| \scK\in\scE_\scK \Big\} }.

The maximally mixed state itself forms a locally scrambled state ensemble $\scE_{\rho_\id}=\{\rho_\id\}$, whose Pauli weight $w_{\scE_{\rho_\id}}$ is given by \eqnref{eq: w maxmix}. Suppose the random channel ensemble $\scE_\scK$ is also locally scrambled, \eqnref{eq: wP evolve} can be applied to compute the Pauli weight of the composed state ensemble $\scE_{\scK(\rho_\id)}$,
\eq{\label{eq: wP evolve2}
w_{\scE_{\scK(\rho_\id)}}(P)=\sum_{P'}w_{\scE_\scK}(P,P')w_{\scE_{\rho_\id}}(P').}
By \eqnref{eq: def Pauli weight}, $w_{\scE_{\scK(\rho_\id)}}$ is defined as
\eq{\label{eq: def Pauli weight2}
w_{\scE_{\scK(\rho_\id)}}=\E_{\scK\in\scE_\scK}(\Tr P\scK(\rho_\id))^2.}

However, what we wish to calculate is the Pauli weight of the classical snapshot ensemble, defined as
\eq{w_{\scE_\sigma}=\E_{\scK\in\scE_\scK}\Big(\Tr \Big(P\frac{\scK(\rho_\id)}{\Tr \scK(\rho_\id)}\Big)\Big)^2=\E_{\scK\in\scE_\scK}\frac{(\Tr P\scK(\rho_\id))^2}{(\Tr \scK(\rho_\id))^2}.}
This ensemble average is difficult to calculate due to the potential correlation between its numerator and denominator. A common strategy is to approximate the average of ratios by the ratio of averages. Using \eqnref{eq: def Pauli weight2}, we have 
\eq{w_{\scE_\sigma}(P)\simeq \frac{\E_{\scK\in\scE_\scK}(\Tr P\scK(\rho_\id))^2}{\E_{\scK\in\scE_\scK}(\Tr \scK(\rho_\id))^2}=\frac{w_{\scE_{\scK(\rho_\id)}}(P)}{w_{\scE_{\scK(\rho_\id)}}(\id)}.}
Such that the Pauli weight of the classical snapshot ensemble can be approximately estimated as the ratio of two Pauli weights that we can compute by the operator dynamics \eqnref{eq: wP evolve2}. The approximation is expected to be asymptotically exact deep in both the volume-law and area-law phases where the correlated fluctuation between the numerator and denominator is suppressed. However, near the measurement-induced entanglement transition, the approximation will become inaccurate in estimating the location of the critical point and its universal properties. Nevertheless, it still provides an overall good picture of the behavior of the Pauli weight across the entanglement transition.

\section{Quantum Statistical Mechanical Picture}\label{app: statmech}

\subsection{Pauli Weight and Entanglement Feature}

Given the prior snapshot ensemble $\scE_\sigma$, we aim to compute the Pauli weight $w_{\scE_\sigma}(P)$, which plays a central role in classical shadow tomography reconstruction and shadow norm estimation. The Pauli weight is defined as
\eq{w_{\scE_\sigma}(P):=\E_{\sigma\in\scE_\sigma}(\Tr P \sigma)^2,}
for any Pauli operator $P=\prod_{i}P_i$ and $P_i\in\{I,X,Y,Z\}$ in qubit systems.
If $\scE_\sigma$ is a locally scrambled ensemble, its 2nd moment if fully captured by the (2nd R\'enyi) entanglement feature, defined as
\eq{W_{\scE_\sigma}(A):=\E_{\sigma\in\scE_\sigma}\Tr(\sigma^{\otimes 2}\mathsf{SWAP}_A),}
where $A$ denotes a subset of totally $N$ qubits in the system, which can also be encoded as a bit string $A\in\{0,1\}^{\times N}$ such that the qubit marked by 1 belongs to the subset. $\mathsf{SWAP}_A$ denotes the swap operator in the region $A$ between the two replicas of $\sigma$. The Pauli weight and the entanglement feature are related by a linear transformation
\eq{w_{\scE_\sigma}(P)|_{\supp P=A}=\Big(-\frac{1}{3}\Big)^{|A|}\sum_{B\subseteq A}(-2)^{|B|}W_{\scE_\sigma}(B),}
where $|A|$ denotes the size (cardinality) of the set $A$.

\begin{figure}[htbp]
\begin{center}
\includegraphics[scale=0.65]{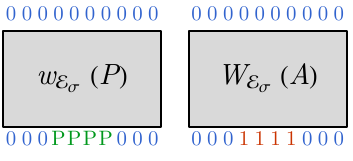}
\caption{Statistical mechanical picture of the Pauli weight $w_{\scE_\sigma}$ and the entanglement feature $W_{\scE_\sigma}$.}
\label{fig: statmech}
\end{center}
\end{figure}

This relation can be simply expressed using the entanglement feature state, which is a fictitious quantum state that encodes the entanglement feature over all possible subsystems $A$. Let $\ket{A}$ be a set of orthonormal bit-string basis (i.e., assuming $\braket{A}{B}=\delta_{AB}$), we can define the entanglement feature state $\ket{W_{\scE_\sigma}}$ for the ensemble $\scE_\sigma$ as
\eq{\ket{W_{\scE_\sigma}}=\sum_{A} W_{\scE_\sigma}(A)\ket{A},}
then both the Pauli weight and the entanglement feature admits simple representation as
\eqs{
w_{\scE_\sigma}(P)&=\braket{P}{W_{\scE_\sigma}},\\
W_{\scE_\sigma}(A)&=\braket{A}{W_{\scE_\sigma}},}
where the state $\ket{P}$ is defined by
\eq{\ket{P}:=\prod_{i\in\supp P}\frac{2X_i-I_i}{3}\ket{\vect{0}},}
with $\ket{\vect{0}}:=\bigotimes_i\ket{0}_i$ being the all-0 state (the basis state of empty region $A=\emptyset$) and $X=\ket{0}\bra{1}+\ket{1}\bra{0}$, $I=\ket{0}\bra{0}+\ket{1}\bra{1}$ are on-site operators in the entanglement feature Hilbert space. Physically, $\ket{0}_i$ (or $\ket{1}_i$) corresponds to the identity (or swap) boundary condition on the qubit $i$ in defining the 2nd-moment computation. The state $\ket{P}$ defines the boundary condition for Pauli weight, corresponding to imposing the Pauli boundary condition within the Pauli operator support.

\subsection{Entanglement Feature State in Measurement-Induced Transition}

Suppose the prior snapshot state is given by $\sigma=K^\dagger K/\Tr(K^\dagger K)$ with $K$ being the Kraus operator that describes a quantum channel of the hybrid quantum circuit with measurement rate $p$.
\begin{itemize}
\item $p=0$ limit, $\scE_\sigma (p=0)$ only contains a maximally mixed state, whose entanglement feature state is
\eq{\ket{W_{\scE_\sigma(p=0)}}=\frac{1}{4^N}W_{\id}\ket{\vect{0}},}
where $W_\id$ is the entanglement feature operator of the identity quantum channel, given by
\eq{\label{eq: Wp=0}
W_\id=2^N\prod_i(2I_i+X_i).}
\item $p=1$ limit, $\scE_\sigma (p=1)$ is an ensemble of random product state, meaning that the entanglement entropy vanishes for all regions, so $\forall A: W_{\scE_\sigma}(A)=1$, the entanglement feature state can be written as
\eq{\label{eq: Wall+}\ket{W_{\scE_\sigma(p=1)}}=2^{N/2}\ket{\vect{+}},}
where $\ket{\vect{+}}:=\bigotimes_i\ket{+}_i$ is the all-plus state with $\ket{+}=\frac{1}{\sqrt{2}}(\ket{0}+\ket{1})$ being the eigenstate of $X=+1$. Given that $\ket{\vect{+}}$ is also an eigenstate of $W_\id$, it does not hurt to rewrite 
\eqnref{eq: Wall+} as
\eq{\label{eq: Wp=1}
\ket{W_{\scE_\sigma(p=1)}}=\Big(\frac{1}{3\sqrt{2}}\Big)^N W_{\id}\ket{\vect{+}}.}
\end{itemize}
Given the two limits \eqnref{eq: Wp=0} and \eqnref{eq: Wp=1}, we can propose a variational ansatz for the entanglement feature state
\eq{\label{eq: WPsi}
\ket{W_{\scE_\sigma(p)}}\propto W_\id\ket{\Psi_p},}
where $\ket{\Psi_p}$ should interpolate between the ferromagnetic state $\ket{\vect{0}}$ and the paramagnetic state $\ket{\vect{+}}$ from $p=0$ to $p=1$. A simple way to realize such an interpolation is to consider $\ket{\Psi_p}$ being the ground state of a transverse field Ising model
\eq{\label{eq: TFIM}
H=-J\sum_{i}Z_{i}Z_{i+1}-h\sum_i X_i,}
where the $h/J$ ratio is supposed to depend on the measurement rate in such a way that
\eq{\left\{
\begin{array}{ll}
h/J=0 & \text{when }p=0,\\
h/J\to\infty & \text{when }p=1.
\end{array}
\right.} 
Within this Ising model description, the measurement-induced entanglement transition happens at $p_c$, corresponding to $h/J=1$ (at the Ising critical point). Although it should be emphasized that this is only a ``mean-field'' description of the entanglement transition, and it is known that the measurement-induced criticality is not in the Ising universality class. Nevertheless, this description provides us with ways to phenomenologically model and describe the Pauli weight for the prior snapshots of hybrid shadow tomography away from the critical point.

Given the variational ansatz of the entanglement feature state $\ket{W_{\scE_\sigma(p)}}$ in \eqnref{eq: WPsi}, the Pauli weight can be evaluated as
\eqs{\label{eq: wpP}
w_{\scE_\sigma(p)}(P)&=\frac{\braket{P}{W_{\scE_\sigma(p)}}}{\braket{\vect{0}}{W_{\scE_\sigma(p)}}}=\frac{\bra{\vect{0}}(\prod_{i\in\supp P}\frac{2X_i-I_i}{3})W_\id\ket{\Psi_p}}{\bra{\vect{0}}W_\id\ket{\Psi_p}}\\
&=\frac{\bra{\vect{0}}\prod_{i\in\supp P}X_i\prod_{i\notin\supp P}(2I_i+X_i)\ket{\Psi_p}}{\bra{\vect{0}}\prod_{i}(2I_i+X_i)\ket{\Psi_p}},}
where the denominator is introduced to normalize the Pauli weight (such that $\forall p: w_{\scE_\sigma(p)}(I)=1$) as the ansatz state $\ket{W_{\scE_\sigma(p)}}$ was unnormalized. Introduce a binary indicator 
\eq{\theta_{i\notin P}:=\left\{
\begin{array}{cc}
0 & i\in\supp P,\\
1 & i\notin\supp P,
\end{array}
\right.} 
\eqnref{eq: wpP} can be written in a more compact form 
\eq{\label{eq: wpP2}
w_{\scE_\sigma(p)}(P)=\frac{\bra{\vect{0}}\prod_{i}(2\theta_{i\notin P}I_i+X_i)\ket{\Psi_p}}{\bra{\vect{0}}\prod_{i}(2I_i+X_i)\ket{\Psi_p}}.
}

\subsection{Volume-Law Phase ($p<p_c$)}

At the $p=0$ limit, the entanglement feature state is given by \eqnref{eq: Wp=0}, corresponding to $\ket{\Psi_{p=0}}=\ket{\vect{0}}$ as the ground state of the Ising model \eqnref{eq: TFIM} at its ferromagnetic fixed point ($h/J=0$).  Given that $\ket{\Psi_{p=0}}=\ket{\vect{0}}$ is a product state, the Pauli weight \eqnref{eq: wpP} can be evaluated at each site independently,
\eq{w_{\scE_\sigma(p=0)}(P)=\prod_{i\in\supp P}\frac{\bra{0}X\ket{0}}{\bra{0}2I+X\ket{0}}=0.}
The Pauli weight vanishes for all $P$ because without any measurement ($p=0$) there is no way to infer any information about any non-trivial Pauli observable.

To move away from this extreme limit, we can turn on the transverse field term $h$ in the Ising model \eqnref{eq: TFIM} and use perturbation theory to estimate the corrected ground state $\ket{\Psi_{p}}$ in the $h/J\ll 1$ regime. To the 1st order of $h/J$, we have
\eq{\ket{\Psi_{p}}\simeq \Big(1+\frac{h}{4J}\sum_{i}X_i\Big)\ket{\vect{0}}\simeq \e^{\frac{h}{4J}\sum_{i}X_i}\ket{\vect{0}}=\prod_{i}\e^{\frac{h}{4J}X_i}\ket{\vect{0}},}
where we exponentiate the leading order correction. With this, the Pauli weight becomes
\eq{w_{\scE_\sigma}(P)=\prod_{i\in\supp P}\frac{\bra{0}X\e^{\frac{h}{4J}X}\ket{0}}{\bra{0}(2I+X)\e^{\frac{h}{4J}X}\ket{0}}=\Big(\frac{1}{1+2 \coth(\frac{h}{4J})}\Big)^{|\supp P|}.}
Assuming $h/J\propto p$ is linear in $p$, the above result implies that the shadow norm should scale with the Pauli operator size $k=|\supp P|$ as
\eq{\Vert P\Vert_{\scE_\sigma}^2=\frac{1}{w_{\scE_\sigma}(P)}=(1+2\coth(c p))^k=\beta^k,}
where $c$ is some unknown coefficient. So the base $\beta=1+2\coth(c p)$ will diverge as $\beta\simeq 1/p$ as the measurement rate $p\to 0$ approaches zero, which is consistent with our numerical result in the main text.

\subsection{Area-Law Phase ($p>p_c$)}

At the $p=1$ limit, the entanglement feature state is given by \eqnref{eq: Wp=1}, corresponding to $\ket{\Psi_{p=1}}=\ket{\vect{+}}$ as the ground state of the Ising model \eqnref{eq: TFIM} at is paramagnetic fixed point ($J=0$). Given that $\ket{\Psi_{p=1}}=\ket{\vect{+}}$ is a project state, the Pauli weight \eqnref{eq: wpP} can be evaluated at each site independently,
\eq{w_{\scE_\sigma(p=1)}(P)=\prod_{i\in\supp P}\frac{\bra{0}X\ket{+}}{\bra{0}2I+X\ket{+}}=\prod_{i\in\supp P}\frac{\braket{0}{+}}{3\braket{0}{+}}=\frac{1}{3^{|\supp P|}},}
given that $X\ket{+}=\ket{+}$. This reproduces the known result of Pauli weight for the random product state ensemble.

To move away from this extreme limit, we can turn on the Ising coupling $J$ in the Ising model \eqnref{eq: TFIM} and use perturbation theory to estimate the corrected ground state $\ket{\Psi_p}$ in the $J/h\ll 1$ regime. To the 1st order of $J/h$, we have
\eq{\ket{\Psi_p}\simeq\Big(1+\frac{J}{4h}\sum_{i}Z_i Z_{i+1}\Big)\ket{\vect{+}}.}
With this, we can evaluate the inner product
\eqs{\label{eq: string}
\bra{\vect{0}}\prod_{i}(2\theta_{i\notin P}I_i+X_i)\ket{\Psi_p}&\simeq \bra{\vect{0}}\prod_{i}(2\theta_{i\notin P}I_i+X_i)\ket{\vect{+}}+\frac{J}{4h}\sum_{j}\bra{\vect{0}}\prod_{i}(2\theta_{i\notin P}I_i+X_i)Z_jZ_{j+1}\ket{\vect{+}}\\
&=\Big(1+\frac{J}{4h}\sum_{j}\frac{2\theta_{j\notin P}-1}{2\theta_{j\notin P}+1}\frac{2\theta_{j+1\notin P}-1}{2\theta_{j+1\notin P}+1}\Big)\prod_{i}(2\theta_{i\notin P}+1)\braket{\vect{0}}{\vect{+}}\\
&\simeq \e^{\frac{J}{4h}\sum_{j}\phi_{j,P}}\prod_{i}(2\theta_{i\notin P}+1)\braket{\vect{0}}{\vect{+}}\\
&= \prod_{i}\e^{\frac{J}{4h}\phi_{i,P}}(2\theta_{i\notin P}+1)\braket{\vect{0}}{\vect{+}},
}
where we have introduced the symbol
\eq{\phi_{i,P}=\frac{2\theta_{j\notin P}-1}{2\theta_{j\notin P}+1}\frac{2\theta_{j+1\notin P}-1}{2\theta_{j+1\notin P}+1}
=\left\{
\begin{array}{ll}
1 & i+\frac{1}{2}\in\supp P,\\
-\frac{1}{3}& i+\frac{1}{2}\in\partial\supp P,\\
\frac{1}{9}& i+\frac{1}{2}\notin\supp P,\\
\end{array}
\right.}
and exponentiate the perturbation (given $J/H\ll 1$). Given the result \eqnref{eq: string}, we can evaluate the Pauli weight using \eqnref{eq: wpP2},
\eqs{w_{\scE_\sigma(p)}(P)&=\frac{\prod_{i}\e^{\frac{J}{4h}\phi_{i,P}}(2\theta_{i\notin P}+1)\braket{\vect{0}}{\vect{+}}}{\prod_{i}(3\e^{\frac{J}{4h}\frac{1}{9}})\braket{\vect{0}}{\vect{+}}}\\
&\simeq \prod_{i\in\supp P}\frac{1}{3}\e^{\frac{J}{4h}\left(1-\frac{1}{9}\right)}\\
&=\Big(\frac{1}{3\e^{-\frac{2J}{9h}}}\Big)^{|\supp P|}.}
Assuming $J/h\propto (1-p)$ is linear in $(1-p)$, the above result implies that the shadow norm should scale with the Pauli operator size $k=|\supp P|$ as
\eq{\Vert P\Vert_{\scE_\sigma}^2=\frac{1}{w_{\scE_\sigma}(P)}=(3\e^{-c'(1-p)})^k=\beta^k,}
where $c'$ is some unknown coefficient. The base $\beta=3\e^{-c'(1-p)}\simeq 3(1-c'(1-p))$ will decrease from $3$ as $p$ deviates from $1$, which is also consistent with our numerical result in the main text.

\subsection{Entanglement Transition ($p=p_c$)}

At $p=p_c$, the hybrid quantum circuit undergoes measurement-induced entanglement transition. Although the transition is not described by the Ising CFT, it is instructive to gain a qualitative understanding about the transition using the Ising analogy. In the Ising analogy, the entanglement transition corresponds to the Ising critical point in the Ising model \eqnref{eq: TFIM}.

Suppose the ground state $\ket{\Psi_p}$ is now described by the Ising CFT. Using the Kramers-Wannier duality: $X_i\leftrightarrow \tilde{Z}_{i-1/2} \tilde{Z}_{i+1/2}$, the string operator $\prod_{i\in A} X_i$ for a contiguous region $A$ can be mapped to the product of the dual Ising operator $\prod_{j\in\partial A}\tilde{Z}_{j}$, such that
\eq{\label{eq: bCFT}
\bra{\vect{0}}\prod_{i\in A}X_i\ket{\Psi_p}=\bra{\vect{0}}\prod_{j\in \partial A}\tilde{Z}_j\ket{\Psi_p}\sim |A|^{-2\Delta}\braket{\vect{0}}{\Psi_p},}
where $\Delta$ corresponds to the scaling dimension of the dual Ising operator in the boundary CFT. Using this interpretation, we can expand
\eqs{\bra{\vect{0}}\prod_{i\in A}(2I_i+X_i)\ket{\Psi_p}
&=2^{|A|}\sum_{B\subseteq A}\frac{1}{2^{|B|}}\bra{\vect{0}}\prod_{i\in B}X_i\ket{\Psi_p}\\
&=2^{|A|}\sum_{B\subseteq A}\frac{1}{2^{|B|}}\bra{\vect{0}}\prod_{j\in \partial B}\tilde{Z}_j\ket{\Psi_p}.
}
We can give a loose bound for the correlation by
\eq{\delta_{B\emptyset}\leq \frac{\bra{\vect{0}}\prod_{j\in \partial B}\tilde{Z}_j\ket{\Psi_p}}{\braket{\vect{0}}{\Psi_p}}\leq 1,}
therefore
\eq{\label{eq: bound}
2^{|A|}=2^{|A|}\sum_{B\subseteq A}\frac{1}{2^{|B|}}\delta_{B\emptyset}\leq \frac{\bra{\vect{0}}\prod_{i\in A}(2I_i+X_i)\ket{\Psi_p}}{\braket{\vect{0}}{\Psi_p}}\leq 2^{|A|}\sum_{B\subseteq A}\frac{1}{2^{|B|}}=2^{|A|}\Big(1+\frac{1}{2}\Big)^{|A|}=3^{|A|}.}
Based on \eqnref{eq: wpP}, the Pauli weight can be estimated by
\eq{w_{\scE_\sigma(p)}(P)\simeq\frac{\bra{\vect{0}}\prod_{i\in\supp P}X_i\ket{\Psi_p}}{\bra{\vect{0}}\prod_{i\in\supp P}(2I_i+X_i)\ket{\Psi_p}},}
which, given \eqnref{eq: bCFT} and \eqnref{eq: bound}, can be bounded by
\eq{2^{-k}k^{-2\Delta}\geq w_{\scE_\sigma(p)}(P)\geq 3^{-k}k^{-2\Delta},}
where $k=|\supp P|$ is the size of the Pauli operator. Thus we conclude that the Pauli weight at $p=p_c$ should take the form of 
\eq{w_{\scE_\sigma(p)}(P)=\beta^{-k}k^{-2\Delta},}
with a base $2\leq \beta\leq 3$. Correspondingly, the shadow norm takes the form of
\eq{\Vert P\Vert_{\scE_\sigma}^2=\frac{1}{w_{\scE_\sigma}(P)}=\beta^k k^{2\Delta},}
a form that is consistent with our numerical result in the main text. Our fitting in the main text shows that $\beta\approx 2.2$ is within the bound.

\subsection{Summary of Results}
In conclusion, the quantum statistical mechanical model presented in this appendix shows that the shadow norm takes the following form
\eq{\Vert P\Vert_{\scE_\sigma}^2=\left\{
\begin{array}{ll}
\beta_\text{vol}^k & p<p_c,\\
\beta_\text{min}^k k^{2\Delta} & p=p_c,\\
\beta_\text{are}^k & p=p_c.\\
\end{array}
\right.}
\begin{itemize}
\item In the volume law phase, the base $\beta_\text{vol}\sim 1/p$ diverges as $p\to 0$.
\item In the area law phase, the base $\beta_\text{are}\sim 3\e^{-c'(1-p)}$ approaches 3 from below as $p\to 1$.
\item Given that $\beta$ increases in both phases as we go away from the critical point, the entanglement transition should have the minimal $\beta$, denoted as $\beta_\text{min}$, and (loosely) bounded by $2\leq \beta_\text{min}\leq 3$.
\item At the critical point, there is a power law correction $k^{2\Delta}$, with universal exponent $\Delta$.
\end{itemize}

\section{Toy Models}\label{app: toy}

\subsection{Area-Law Phase ($p>p_c$)}

\begin{figure}[htbp]
\begin{center}
\includegraphics[scale=0.65]{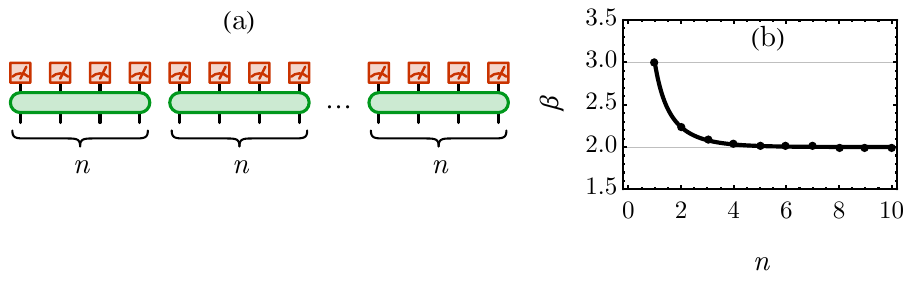}
\caption{(a) A toy model for the randomized measurement in the area-law phase. Each green block represents a $n$-qubit random Clifford gate. (b) The dependence of the shadow norm scaling base $\beta$ on the block size $n$.}
\label{fig: toy area}
\end{center}
\end{figure}

When the measurement rate $p$ is greater than the critical value $p_c$, the classical snapshot $\sigma$ is in the area-law phase. The measurement circuit can be modeled by \figref{fig: toy area}(a), whose corresponding classical snapshot state $\sigma$ is a product of $n$-qubit random stabilizer state. Within each block, the entanglement is maximal. Between different blocks, there is no entanglement (product state). Therefore, the block size $n$ parameterizes the typical range of local entanglement in the classical snapshot state $\sigma$. 

The limit $p=1$ corresponds to $n=1$, that every qubit gets measured immediately before any inter-qubit scrambling. As the measurement rate $p$ reduces (but still in the area-law phase $p_c<p<1$), the system will be scrambled within the local entanglement range before gets probed by the measurement, but the scrambling range is still finite. This situation can be modeled by $n>1$. We expect $n$ to increase effectively as we $p$ reduces from 1.

Consider a Pauli observable $P$ whose support happens to cover $m$ of the $n$-qubit blocks. On one hand, the operator size of $P$ is $|\supp P|=mn$. On the other hand, its Pauli weight is given by
\eq{\label{eq: wP area}
w_{\scE_\sigma}(P)=\Big(\frac{2^n-1}{4^n-1}\Big)^m=\frac{1}{(2^n+1)^m}.}
This result can be understood as follows. The Pauli weight can be interpreted as the probability that a Pauli observable $P$ gets transformed by the random Clifford gates into a diagonal operator in the measurement basis, such that it can be directly probed by the measurement. Within each $n$-qubit block, the random Clifford gate scrambles any particular non-identity Pauli observable to one of all $(4^n-1)$ non-identity Pauli observables, among which only $(2^n-1)$ are diagonal (as Pauli strings of $I$ and $Z$) in the measurement basis. Hence the probability of $P$ to be diagonalized within each block is $(2^n-1)/(4^n-1)=1/(2^n+1)$. To diagonalize $P$ across all the $m$ blocks, the probability multiplies to $1/(2^n+1)^m$, as concluded in \eqnref{eq: wP area}. As a result, the shadow norm of $P$ is
\eq{\Vert P\Vert_{\scE_\sigma}^2=\frac{1}{w_{\scE_\sigma}(P)}=(2^n+1)^m.}
Assuming the shadow norm scales with the operator size as $\Vert P\Vert_{\scE_\sigma}^2=\beta^{|\supp P|}$, the base $\beta$ can be extracted as
\eq{\log\beta=\lim_{|\supp P|\to \infty}\frac{\log \Vert P\Vert_{\scE_\sigma}^2}{|\supp P|}=\lim_{m\to\infty}\frac{m\log(2^n+1)}{mn}=\frac{\log(2^n+1)}{n}.}
The dependence of $\beta$ on $n$ is shown in \figref{fig: toy area}(b). As the measurement rate $p$ decreases from 1, the scrambling range $n$ increases, and $\beta$ will eventually decrease from $3$ to $2$ (in this toy model's ideal case). 

\subsection{Volume-Law Phase ($p<p_c$)}

\begin{figure}[htbp]
\begin{center}
\includegraphics[scale=0.65]{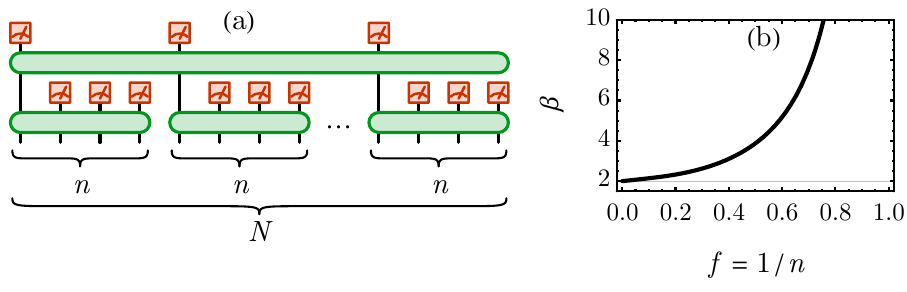}
\caption{(a) A toy model for the randomized measurement in the volume-law phase. Each green block represents a random Clifford gate. The lower layer gates serve as the encoding gates of random stabilizer codes in each $n$-qubit block. The upper layer gate scrambles all the $N/n$ logical bits. (b) The dependence of the shadow norm scaling base $\beta$ on the volume-law coefficient $f$. }
\label{fig: toy volume}
\end{center}
\end{figure}

When the measurement rate $p$ is less than the critical value $p_c$, the classical snapshot $\sigma$ is in the volume-law phase with local error correction encoding. A model to describe such states is to consider random stabilizer states further encoded by random stabilizer codes. The corresponding measurement circuit will be like \figref{fig: toy volume}(a). The system contains $N$ physical qubits, grouped by $n$-qubit blocks. Within each block, the $n$ physical qubit is first decoded into a logical qubit and $(n-1)$ syndrome qubits. The syndrome qubits are measured, and the logical qubits are further scrambled before being finally measured. We assume $N\gg n\gg 1$.

Consider an entanglement region $A$ that covers exactly $m$ of the $n$-qubit blocks (assuming $mn\ll N$), the entanglement entropy of $\sigma$ in such region scales as $S_\sigma(A)=m \log 2$, while the region size is $|A|=mn$. So $\sigma$ is a volume-law state with the volume-law coefficient
\eq{\label{eq: vlc}
f=\frac{S_\sigma(A)}{|A|\log 2}=\frac{1}{n}.}
This $1/n$ ratio is also the rate of the local error correction code (the ratio between the numbers of logical v.s. physical qubits). As the measurement rate $p$ increases from 0 to $p_c$, the volume-law coefficient $f$ decreases from 1 to 0, corresponding to $n$ increasing from 1 to $\infty$. 

Consider a Pauli observable $P$ whose support happens to cover $m$ of the $n$-qubit blocks. On one hand, the operator size of $P$ is $|\supp P|=mn$. On the other hand, its Pauli weight is given by
\eqs{\label{eq: wP volume}
w_{\scE_\sigma}(P)&=q^m + \varepsilon (r^m-q^m),\quad\text{with }\\
\varepsilon &=\frac{2^{N/n}-1}{4^{N/n}-1}=\frac{1}{2^{N/n}+1},\\
q&=\frac{2^{n-1}-1}{4^n-1},\\
r&=\frac{4\times 2^{n-1}-1}{4^n-1}.}
This result can be understood as follows. The Pauli weight can be interpreted as the probability that a Pauli observable $P$ gets transformed by the random Clifford gates into a diagonal operator in the measurement basis, such that it can be directly probed by the measurement. There are two possible scenarios:
\begin{itemize}
\item $P$ gets transformed to an operator that is non-identity and diagonal in the syndrome subspace and identity in the logical subspace. The probability for this to happen is $q^m$. Because in each block, any particular non-identity Pauli observable will be transformed to one of the $(4^n-1)$ non-identity Pauli observable, among which only $(2^{n-1}-1)$ are syndrome subspace diagonal and logical subspace identity. They are of the following form:
\eq{I\otimes \mat{I\\Z}^{\otimes (n-1)}\text{ excluding }I^{\otimes n}.}
So the probability for this to happen is $q=(2^{n-1}-1)/(4^n-1)$ in each block and $q^m$ over $m$ blocks. In this case, $P$ will be directly measured by the syndrome qubit measurements. After the measurement, it collapses to an identity operator on logical qubits, whose measurement outcome can be determined with probability 1. Therefore, this scenario has a total contribution of $q^m\times 1$ to the Pauli weight, corresponding to the first term in \eqnref{eq: wP volume}.

\item $P$ gets transformed to an operator that is non-identity and diagonal in both the syndrome and logical subspaces. The probability for this to happen is $\varepsilon (r^m-q^m)$. Within each block, any particular non-identity Pauli observable will be transformed to one of the $(4^n-1)$ non-identity Pauli observables, among which only $(4\times 2^{n-1}-1)$ are diagonal in the syndrome subspace regardless of its action in the logical subspace. They are of the following form
\eq{\mat{I\\X\\Y\\Z}\otimes\mat{I\\Z}^{\otimes(n-1)}\text{ excluding }I^{\otimes n}.}
So the probability for this to happen is $r=(4\times 2^{n-1}-1)/(4^n-1)$ and $r^m$ over $m$ blocks. After the syndrome qubit measurements, the operator will collapse to  a Pauli string in the logical subspace of the following form
\eq{\mat{I\\X\\Y\\Z}^{\otimes m}.}
This has not excluded the possibility of $I^{\otimes m}$, which should be excluded to avoid double counting the same scenario discussed previously. Thankfully, we already know that the probability for $P$ to become a logical identity operator after the syndrome qubit measurement is $q^m$. So the probability for $P$ to be syndrome diagonal and logical non-identity is $(r^m-q^m)$. Under the global scrambling of $N/n$ logical qubits by the upper layer gate, the observable $P$ gets further mapped to one of the $(4^{N/n}-1)$ non-identity Pauli operators in the logical subspace, among which only $(2^{N/n}-1)$ are diagonal and can be directly probed by the logical qubit measurements. So this further multiplies the probability by $\varepsilon=(2^{N/n}-1)/(4^{N/n}-1)$, resulting in a contribution of $\varepsilon(r^m-q^m)$ to the Pauli weight, corresponding to the second term in \eqnref{eq: wP volume}.
\end{itemize}
According to \eqnref{eq: wP volume}, $\varepsilon\to 0$ in the thermodynamic limit $N\gg n$, so the second term can be neglected, and the Pauli weight is dominated by the first term
\eq{
w_{\scE_\sigma}(P)\overset{N\gg n}=q^m=\Big(\frac{2^{n-1}-1}{4^n-1}\Big)^m.}

In conclusion, the shadow norm of $P$ in this toy model is given by
\eq{\Vert P\Vert_{\scE_\sigma}^2=\frac{1}{w_{\scE_\sigma}(P)}\overset{N\gg n}=\Big(\frac{4^n-1}{2^{n-1}-1}\Big)^m.}
Assuming the shadow norm scales with the operator size as $\Vert P\Vert_{\scE_\sigma}^2=\beta^{|\supp P|}$, the base $\beta$ can be extracted as
\eq{\log\beta=\lim_{|\supp P|\to \infty}\frac{\log \Vert P\Vert_{\scE_\sigma}^2}{|\supp P|}=\lim_{m\to\infty}\frac{1}{mn}m\log\Big(\frac{4^n-1}{2^{n-1}-1}\Big)=\frac{1}{n}\log\Big(\frac{4^n-1}{2^{n-1}-1}\Big).}
Given that $n$ is related to the volume law coefficient $f=1/n$ by \eqnref{eq: vlc}, the base $\beta$ can also be expressed in terms of $f$ as
\eq{\log\beta=f\log\Big(\frac{4^{1/f}-1}{2^{1/f-1}-1}\Big)\overset{f\to 0}=(1+f)\log 2.}
The dependence of $\beta$ on $f$ is shown in \figref{fig: toy volume}(b). As the measurement strength increases from $0$ to $p_c$, the volume-law coefficient $f$ decreases from $1$ to $0$, and $\beta$ decreases from $\infty$ to $2$. Near the measurement induced phase transition (as $f\to 0$), $\beta$ increases with the volume-law coefficient $f$ as $\beta= 2^{1+f}$.

\end{document}